\newif\iffigs\figstrue

\iffigs
   \documentstyle[12pt,epsf]{article}
\else
   \documentstyle[12pt]{article}
\fi

\newcommand{\eq}{\begin{equation}}
\newcommand{\en}{\end{equation}}
\newcommand{\eqa}{\begin{eqnarray}}
\newcommand{\ena}{\end{eqnarray}}

\def\ee#1{{\rm e}^{#1}}
\def\trace{{\rm Tr}\hskip 1pt}
\def\ii{{\rm i}}

\def\ibes#1{{I_{#1}(\beta_t)\over I_1(\beta_t)}}
\def\ibessp#1{{I_{#1}(\beta_s)\over I_1(\beta_s)}}
\def\ibe#1#2{\left[{I_{#1}\over I_1}\right]^{#2}}

\def\cj{{\cal C}^{(j)}}
\def\vx{{\vec x}}
\def\cmn#1#2{{\cal C}^{(m,n)}_{#1#2}}
\def\um{{1\over 2}}

\def\ctmn{{\widetilde{\cal C}}^{(m,n)}}
\def\czmn{{\cal C}^{(m,n)}_0}
\def\cumn{{\cal C}^{(m,n)}_1}
\def\bpmn{{\cal B}^{(m,n)}_+}
\def\bmmn{{\cal B}^{(m,n)}_-}
\def\cz#1#2{{\cal C}_0^{(#1,#2)}}
\def\cu#1#2{{\cal C}_1^{(#1,#2)}}

\def\bpm#1#2{{\cal B}_\pm^{(#1,#2)}}

\newcommand{\NP}[1]{Nucl.\ Phys.\ {\bf #1}}
\newcommand{\PL}[1]{Phys.\ Lett.\ {\bf #1}}

\newcommand{\PR}[1]{Phys.\ Rev.\ {\bf #1}}
\newcommand{\PRL}[1]{Phys.\ Rev.\ Lett.\ {\bf #1}}

\begin{document}

\hskip 10.5cm \vbox{\hbox{tt hep-lat/9601020}
\hbox{DFTT 69/95}\hbox{NORDITA 96/1P}
\hbox{January 1996}}
\vskip 0.4cm
\centerline{\Large\bf Toward an analytic determination of the}
\centerline{\Large\bf deconfinement temperature in SU(2) L.G.T.}
\vskip 1cm
\centerline{M. Bill\'o$^{a,b}$ \footnote{billo@nbi.dk},M. Caselle
$^{b,c}$
\footnote{caselle@to.infn.it}, A. D'Adda$^{b,c}$, and S. Panzeri$^{d}$}
\vskip .5cm
\centerline{\sl $^a$ NORDITA, Blegdamsvej 17, K\o\hskip 1pt benhavn
\O, Denmark}
\vskip .1cm
\centerline{\sl $^b$ Istituto Nazionale di Fisica Nucleare, Sezione di Torino}
\vskip .1cm
\centerline{\sl  $^c$ Dipartimento di Fisica
Teorica dell'Universit\`a di Torino}
\centerline{\sl via P.Giuria 1, I-10125 Turin,Italy}
\vskip .1cm 
\centerline{\sl $^d$  SISSA, Via Beirut 2-4, 34013 Trieste, Italy}

\vskip 1.5cm

\begin{abstract}
We consider the SU(2) lattice gauge theory at finite temperature in ($d$+1)
dimensions, with different couplings $\beta_t$ and $\beta_s$ for timelike and
spacelike plaquettes. By using the character expansion of the Wilson action and
performing the integrals over space-like link variables, we find an effective
action for the Polyakov loops which is {\sl exact to all orders in $\beta_t$}
and to the first non-trivial order in $\beta_s$. The critical coupling for the
deconfinement transition is  determined in the (3+1) dimensional case, 
by the mean field method, for different values of the lattice size $N_t$ in the
compactified time direction and of the asymmetry parameter $\rho = 
\sqrt{\beta_t/\beta_s}$.  We find good agreement with Montecarlo
simulations in the range $1\leq N_t \leq 5$, and good qualitative agreement in
the same range with the logarithmic scaling law of QCD. 
Moreover the dependence of the results from the parameter $\rho$ is in
excellent agreement with previous theoretical predictions.

\end{abstract}

\vfill \eject

\newpage

\setcounter{footnote}{0}
\def\thefootnote{\arabic{footnote}}

\section{Introduction}

The aim of this paper is to obtain, by using only analytical
methods, reliable estimates of the deconfinement temperature in the
SU(2) pure gauge theory (namely without quarks) in (3+1)
dimensions.
The natural framework to pose this question is that of the finite
temperature Lattice Gauge Theories (LGT).  In this framework,
during these last years, the best estimates of the
deconfinement temperature have been obtained by means of Montecarlo simulations,
which are certainly the most powerful tool to extract quantitative results from
LGT. However we think that it is important in itself
to have some independent analytical estimate of the
location of the critical point, besides the outputs of the computer
simulations, to reach a deeper theoretical understanding of the deconfinement
transition.
The attempts to obtain analytically the critical
temperature have a rather long history, starting  more than ten years
ago~\cite{og,djk,gk,gw,ps}.
However the strategy has always been essentially the same: first, construct an
effective action in terms of the Polyakov loops (which, as we shall see below,
 are the relevant
dynamical variables in the physics of deconfinement for pure gauge theories).
Second, use a mean field approximation to extract the critical coupling.
A common feature of all these attempts was that
the effective actions were always
 constructed neglecting the spacelike part of the
action. As a consequence it was impossible to reach
a consistent continuum limit for the critical temperature.

The aim of this paper is to show that it is possible
to overcome this problem. We shall construct in the $SU(2)$
case an improved effective action which takes into account also the spacelike
part of the original Wilson action and is {\sl exact to all orders in the timelike
coupling}.
This is a rather non trivial result and
we shall devote most of this paper to describe how it can be obtained.
Moreover, as we shall see, our approach is a  constructive one and can be
extended in principle to all orders in the space-like couplings.

We decided in this paper to concentrate
only on the gauge group  $SU(2)$ for simplicity reasons, but most of our
results can be  extended to $SU(N)$ models with $N>2$.
Indeed, this
paper can be considered as the natural continuation of~\cite{bcdmp}
where these same techniques were applied to the $N \to \infty$ limit of
LGT. Here we try to eliminate the large $N$
 approximation by looking directly at
the $N=2$ case.

This paper is organized as follows: after a short introduction to
finite temperature lattice gauge theory (sect. 2), we shall devote
sect. 3 to the construction of the  effective action.
In particular, sections \ref{adjoint} and \ref{fundamental} contain
the computation of the first non-trivial contributions from the space-like
part of the action; these sections are rather technical and the reader
interested mainly in the results may wish to skip them, as the results are
anyhow summarized in section \ref{summary}. 
In sect. 4 we shall extract the critical
deconfinement temperature with  mean field 
techniques, we shall discuss our results in comparison with  existing
Montecarlo estimates and check their consistency in the case of asymmetric
lattices with known theoretical results. 
Finally sect. 5 will be devoted to some concluding remarks.
We shall try to keep our formalism as general as possible, so we shall
derive in sect. 2 and 3 the effective action for the Polyakov loop
in a $(d+1)$-dimensional LGT with an arbitrary $d$, and we
shall fix $d=3$ only in  sect.~4.

\section{Finite Temperature LGT}

\subsection{General Setting}

Let us consider a pure gauge theory with gauge group $SU(2)$, defined
on a $d+1$ dimensional cubic lattice.
In order to describe a finite temperature LGT,
we have to impose periodic
boundary conditions in one direction (which we shall call from now
on ``time-like'' direction), while the boundary conditions in the
other $d$ direction (which we shall call ``space-like'') can be
chosen freely.
We take a lattice of $N_t$ ($N_s$) spacings in the time (space)
direction, and we work with the pure gauge theory, containing
only gauge fields described by the link
variables $U_{n;i} \in SU(2)$, where $ n \equiv (\vec x,t)$
denotes the space-time position of the link and $i$ its direction.
It is useful to choose different bare couplings in the time and space
directions. Let us call them $\beta_t$ and $\beta_s$ respectively.
The Wilson action is then
\eq
S_W=\sum_{n}~\frac{1}{2}\left\{\beta_t\sum_i~{\rm Tr_f}(U_{n;0i})
+\beta_s\sum_{i<j}~{\rm Tr_f}(U_{n;ij})\right\}~~~,
\label{wilson}
\en
where ${\rm Tr_f}$ denotes the trace in the fundamental representation
and $U_{n;0i}$ ($U_{n;ij}$) are the time-like (space-like)
plaquette variables, defined as usual by
\eq
U_{n;ij}=U_{n;i}U_{n + i;j}
U^\dagger_{n + j;i}U^\dagger_{n;j}~~~.
\en

In the following we shall call $S_s$ ($S_t$) the space-like (time-like)
part of $S_W$.

Let us  introduce an asymmetry parameter $\rho$ defined by the relation:
 $\beta_t/\beta_s\equiv\rho^2$. 
As $\rho$ varies we have
different, but equivalent, lattice regularization of the same model. This
equivalence is summarized by the following equations, which
can be obtained by taking the classical continuum limit of (\ref{wilson}) and 
which relate $\beta_s$ and $\beta_t$  to the (bare) gauge
coupling $g$ and to the temperature $T$:
\eq
\frac{4}{g^2}=a^{3-d}\sqrt{\beta_s\beta_t}~~~,
\hskip 1cm
T=\frac{1}{N_ta}\sqrt{\frac{\beta_t}{\beta_s}}~~~.
\label{couplings}
\en
Here $a$ is the space-like lattice spacing and  $\frac{1}{N_tT}$ is
the time-like spacing, hence $\rho$ is the ratio between the two. 
{}From this last observation it is clear that equivalent regularizations
with different values of $\rho$ require different
values of $N_t$. Hence, to maintain the  equivalence, $N_t$ must be
a function of $\rho$: $N_t(\rho)$. 

Among all these equivalent regularizations a particular role is played by the
symmetric one, which is defined by:
\eq
\beta\equiv\frac{4}{g^2} a^{d-3}
\label{coupsym}
\en
(from now on we shall distinguish the symmetric regularization from the
asymmetric ones by eliminating the subscripts $t$ and $s$ in $\beta$).
Comparing eqs.(\ref{couplings},\ref{coupsym}) we see that all the
regularizations are equivalent if the following relations hold:
\eq
\beta=\rho\beta_s=\frac{\beta_t}{\rho} ,
\label{rel1}
\en
\eq
N_t(\rho)=\rho N_t(\rho=1) .
\label{rel2}
\en
Notice however that these equivalence relations are constructed in the 
{\sl naive} or ``classical'' continuum limit. At the quantum level, in the (3+1)
dimensional case these
relations change  slightly. These modifications have been studied by F.Karsch
in~\cite{k81} and we shall discuss them in sect.4 . Let us now only anticipate
that eq.(\ref{rel1}) becomes:
\eq
\beta_t=\rho(\beta+4c_\tau(\rho))
\label{rel3}
\en
\eq
\beta_s=\frac{\beta+4c_\sigma(\rho)}{\rho} ,
\label{rel4}
\en
where the two functions $c_\sigma(\rho)$ and $c_\tau(\rho)$ can be found 
in~\cite{k81}
\footnote{We have chosen, for sake of clarity, to keep the notation 
of~\cite{k81} for $c_\sigma$ and $c_\tau$.
This explains the factor 4 in the above equations}. For our purposes we only
need to know the first terms of
 their $1/\rho$ expansion in the $\rho\to\infty$ limit. Let us define:
\eq
4 c_{\sigma,\tau}\equiv \alpha^{0}_{\sigma,\tau}
+\frac{\alpha^{1}_{\sigma,\tau}}{\rho}+\cdots .
\label{rhoex}
\en

The $\alpha$'s can be calculated from the expression of $c_{\sigma,\tau}(\rho)$
given in~\cite{k81}; they are:
$\alpha^{0}_{\tau}=-0.27192$; $\alpha^{1}_{\tau}=1/2$;
$\alpha^{0}_{\sigma}=0.39832$; $\alpha^{1}_{\sigma}=0$. 

These relations will play a major role in the following, since the invariance
of our results as $\rho$ changes is a crucial consistency
check of all our approach. Actually it is a  non trivial
test, since these relations are the result of a {\sl weak coupling
calculation}, and only become manifest in the continuum limit of the model.
Being able to reproduce  them within the framework of a strong coupling 
calculation would be a remarkable and a priori unexpected result.
This is actually the case, as discussed in detail in sect. 4.

A second reason for which it is important to have under control this $\rho$
symmetry is that, at the end, we would like to compare our prediction with the
Montecarlo simulations, which are all made on symmetric lattices. However, at
the same time, it is only in the limit of highly asymmetric lattices ($\rho
\to\infty$) that (as we shall see in more detail below) we can define in an
unambiguous way our expansion of the spacelike part of the action, so it is
somehow mandatory for us to be able to match these two different limits. 

\vskip 0.2cm

In a finite temperature discretization it is possible to define
gauge invariant observables which are topologically non-trivial,
as a consequence of the periodic boundary conditions in the
time directions.
The simplest choice is the Polyakov loop 
defined  in terms of link variables as:
\eq
\hat P_{\vec x}\equiv {\rm Tr}
 \prod_{t=1}^{N_t}(V_{\vec x,t})~~~.
\label{polya}
\en
where $V_{\vec x,t} \equiv U_{\vec x, t ; 0}$ are the vertical link matrices.
In the following we shall often use the untraced quantity $P_{\vec x}$, defined
as:
\eq
 P_{\vec x}\equiv
 \prod_{t=1}^{N_t}(V_{\vec x,t})~~~.
\label{polya2}
\en
which will be referred to as ``Polyakov line''.

As it is well known, the finite temperature theory has a new
global symmetry (unrelated to the gauge symmetry), with symmetry group
the center $C$ of the gauge group (in our case $Z_2$). The Polyakov loop
is a natural order parameter for this symmetry.

In $d>1$, finite temperature gauge theories admit a
deconfinement transition at $T=T_c$, separating the
high temperature, deconfined, phase ($T>T_c$) from the low
temperature, confining domain ($T<T_c$).
In the following we shall be interested in the phase diagram of the
model as a function of $T$, and we shall make some attempt to locate the
critical point $T_c$. The high
temperature regime is characterized by the  breaking of the global
symmetry with respect to the center of the group. In this phase the
Polyakov loop has a non-zero expectation value, and it is
an element of the center of the gauge group
(see for instance~\cite{sy}).

\subsection{Svetitsky-Yaffe conjecture}

The Svetitsky-Yaffe conjecture \cite{sy}
 is based on the idea that, if one were
able to integrate out all the
gauge degrees of freedom of the original $(d+1)$--dimensional
 model {\sl except those related to the Polyakov loops},
then the resulting
effective theory for the Polyakov loops would be a $d$-dimensional spin
system with symmetry group $C$. The deconfinement transition of the
original model would become the order--disorder transition of the
effective spin system.
This effective theory would obviously have very complicate
interactions, but Svetitsky and Yaffe were able to argue that all these
interactions should be short ranged. As a consequence, if the
transition point of the
effective spin system is of second order, near this critical point,
where the correlation length becomes infinite, the precise form of
the short ranged
interactions should not be important, and the universality class of the
deconfinement transition should coincide with that of the simple spin
model with only nearest neighbour
interactions and the same global symmetry group. In particular the
deconfinement transition of the $d+1$
dimensional SU(2) LGT in which we are interested should belong
to the same
universality class of the magnetization transition of the
$d$-dimensional spin Ising model. Unfortunately this argument cannot help to
fix the critical temperature, which is not an universal result, but depends on
the precise form of the action that we study, and hence of the short ranged
interactions that we neglected above. In the next section we shall construct
these correction terms explicitly.

\subsection{Character expansion}

An important role in the following analysis will be played by the
character expansion, which in the SU(2) case is very easy to handle.
Let us briefly summarize few results.
The character of the group element $U$ in the $j^{th}$ representation
is:
\eq
\chi_j(U)\equiv {\rm Tr}_j(U)=\frac{\sin((2j+1)\theta)}{\sin(\theta)}
\label{c1}
\en
where ${\rm Tr_j}$ denotes the trace in the $j^{th}$ representation and
$\theta$ is defined according to the following parametrization of $U$ in
the fundamental representation:
\eq
U=\cos(\theta){\bf 1} + \ii\,\vec\sigma\cdot\vec n \sin(\theta) .
\label{deftheta}
\en
Here $\vec n$ is a tridimensional unit vector and $\sigma_i$ are the
three Pauli matrices. Notice, as a side remark, that with this
parametrization the Haar measure has the following form:
\eq
DU=\sin^2(\theta)\frac{d\theta d^2\vec n}{4\pi^2}
\en
and the Polyakov loop becomes $\hat P_{\vec x}=2\cos(\theta_{\vec x})$

The following orthogonality relations between characters hold:
\eq
\int DU \chi_r(U)~\chi_s(U)=\delta_{r,s}
\label{c2}
\en

\eq
\sum_{r} d_r \chi_r(U~V^{-1})= \delta(U,V) 
\label{c3}
\en

where $d_r$ denotes the dimensions of the $r^{th}$ representation:
$d_r=2r+1$.
In the following we shall use two important properties of the
characters:

\eq
\int DU \chi_r(U)~\chi_s(U^{-1}V)=\delta_{r,s}\frac{\chi_r(V)}{d_r}
\label{c4}
\en
\eq
\int DU \chi_r(UV_1U^{-1}V_2)=\frac{1}{d_r}\chi_r(V_1)\chi_r(V_2) .
\label{c5}
\en

The character expansion of the Wilson action has a particularly simple
form:
\eq
e^{\frac{\beta}{2}{\rm Tr_f}(U)}=\sum_j 2(2j+1)\frac{I_{2j+1}(\beta)}
{\beta}\chi_j(U),~~~j=0,\frac{1}{2},1\cdots
\label{wilsonexp}
\en
where $I_n(\beta)$ is the $n^{th}$ modified Bessel function.
It is customary to collect in front of expression (\ref{wilsonexp}) a factor
of ${I_1(\beta) \over \beta/2}$, so that the expansion starts with 1.

\section{Construction of the Effective Action}
In this section our goal is to construct an effective action for the
finite temperature LGT in terms of the Polyakov loops only. To do so one
should be able  to integrate exactly on the
spacelike variables so that the only remaining degrees of
freedom at the end are the Polyakov loops. Notice that in this
way the resulting effective action would live in $d$ dimensions
(one dimension less than the starting model). This is exactly along the
line of the original Svetitsky-Yaffe program.
As already remarked in the introduction the early attempts to determine
analytically the critical temperature were all based on the assumption
that the deconfinement transition is dominated by the timelike plaquettes,
and the contribution of the space-like plaquettes was 
consequently neglected. Although this approximation correctly predicts
the existence of the deconfinement transition, a quantitative estimate
of the critical temperature for large enough values of $N_t$, namely near
the continuum limit, requires the contribution of the space-like plaquettes
to be taken into account.
Accordingly, we shall treat the timelike part 
of the Wilson action $S_t$ as a Born term and
treat the spacelike part $S_s$
 as a perturbation; namely, we shall make a strong
coupling expansion in $\beta_s$, while
the time-like part of action will be
treated exactly.
This means that order by order in $\beta_s$
the dependence of the effective action  from
$\beta_t$ will be exact, the only expansion parameter being
thus $\beta_s$.
Of course, the zeroth order in $\beta_s$
will contain the timelike plaquettes only. It is not at all obvious
that the integration over the spacelike links could be done to all
orders in $\beta_t$, but it turns out to be the case in the
framework of the characters expansion (see below)  {\sl order by order
in $\beta_s$}.
Rather than a straightforward expansion in powers of $\beta_s$ we 
shall use for each space-like plaquette a character expansion.
Each representation $j$ in the expansion gives a contribution proportional
to a ratio of Bessel functions which is of order $\beta_s^{2 j}$,
so that the character expansion and the expansion in powers
of $\beta_s$ coincide up to higher order terms arising from
the power series expansion of the Bessel functions. 
As it will be discussed later in sect. 4
these higher order terms vanish anyway in the limit of highly
asymmetric lattice. 
We shall consider in this paper only the zeroth order and the
first non trivial order in $\beta_s$, namely $\beta_s^2$.
Terms of order $\beta_s^2$ come either from one space-like plaquette in
the adjoint representation or from a couple of plaquettes in the fundamental 
representation. However it should be noted that there is  no obstruction
in principle to go to higher orders.

For any given order in $\beta_s $ the result is given by an infinite
sum of characters.
Remarkably enough in the $N_t=1$ case this series can be summed exactly
and the result can be written in a closed form. This is essentially due
to the fact the if $N_t=1$ the same effective action can be obtained
in a completely different way, using techniques typical of matrix models
(see below), thus allowing a non trivial  check of all our strong
coupling results.

\subsection{Expansion in $\beta_s$ of the effective action}

The effective action $S_{\rm eff}$ for the Polyakov lines $P_{\vec x}\equiv
\prod_{t=1}^{N_t} V_{\vec x}$ is obtained by integrating over all the
spacelike degrees of freedom in the action (\ref{wilson}). As explained
previously, our approach is to consider the contributions from the spacelike
plaquettes up to a certain order in $\beta_s$ only.
So, for our purposes, it will be convenient to expand separately the spacelike
and the timelike part of the action (\ref{wilson}):
\eqa
\label{gen1}
\exp(S_{\rm eff}) & = & \int\prod_{\vec x,t;i} DU_{\vec x,t;i}
 \exp S_W
  \nonumber \\
& = & \int\prod_{\vec x^\prime,t^\prime;i^\prime} 
DU_{\vec x^\prime,t^\prime;i^\prime} \hskip 2pt
\prod_{\vec x^{\prime\prime}, t^{\prime\prime} ; i^{\prime\prime}} 
\left(1 + \sum_{j={1\over2}}^\infty d_j
\ibes{2j+1} \chi_j(U_{\vec x^{\prime\prime}, t^{\prime\prime};0i^{\prime\prime}}) 
\right)\nonumber\\
& & \times \prod_{\vec x,t; i<j} \left(1 + \sum_{l={1\over2}}^\infty d_l
\ibessp{2l+1} \chi_l(U_{\vec x,t; ij})\right) .
\ena
Specifically, we work out here the effective action
up to $O(\beta_s^2)$. This means that in eq.(\ref{gen1}) we must look only at
the terms containing at most a single space-like plaquette in the adjoint
representation, $\chi_1(U_{\vec x,t;ij})$, or two space-like plaquettes in the
fundamental, $\chi_{1\over 2}(U_{\vec x,t_1;ij}) \chi_{1\over 2}
(U_{\vec y,t_2;kl})$.
Due to the orthogonality relations for characters, it easy to convince
oneself that a pair of plaquettes in the fundamental representation
do actually contribute to the integral only if they appear in the
same spatial position (at two different times $t_1$ and $t_2$);
for the same reason a single fundamental plaquette cannot contribute.
We are thus lead to the following expression:
\eqa
\label{gen2}
&&\exp(S_{\rm eff}) = \int\prod_{\vec x^\prime,t^\prime;i^\prime} 
DU_{\vec x^\prime,t^\prime;i^\prime}\hskip 2pt
\prod_{\vec x^{\prime\prime},t^{\prime\prime}; i^{\prime\prime}} 
\left(1 + \sum_{j={1\over2}}^\infty d_j
\ibes{2j+1} \chi_j(U_{\vec x^{\prime\prime}, t^{\prime\prime}; 0i^{\prime\prime}} ) 
\right)\nonumber \\
&&\times \left( 1 + \sum_{\vec x, i < j} \left[\sum_{t=1}^{N_t} 3 \ibessp{3}
\chi_1(U_{\vec x, t; ij}) + \sum_{t_1 < t_2}
4 \left(\ibessp{2}\right)^2
\chi_{1\over 2}(U_{\vec x, t_1; ij})
\chi_{1\over 2}(U_{\vec x, t_2; ij})\right]\right)
.\nonumber\\
\ena
In the next sections we will consider separately
the three contributions appearing in
(\ref{gen2}). The first one that we will consider corresponds to the
``$1$'' in the second factor above, and gives the $O(\beta_s^0)$ result.

\subsection{Zeroth order approximation}

In the zeroth order approximation we have to consider only
the timelike part of the Wilson action:
\eq
\label{zero1}
\exp(S_0) = \int\prod_{\vec x,i;t}\hskip 2pt\left[DU_{\vec x,i;y}\hskip 2pt
\left(1+\sum_{j={1\over 2}}^\infty d_j\ibes{2j+1} \chi_j(U_{\vec x,i;t}
V_{\vec x + i;t} U_{\vec x,t+1;i}^\dagger V_{\vec x;t}^\dagger)
\right)\right] .
\en
In this case we can easily integrate all the spacelike
links. The reason is that each spacelike link only belongs to two
timelike plaquettes; hence, by making a character expansion, it can be
exactly integrated out. Let us do this integration in two steps, for
future commodity. First let us integrate all the spacelike links except
the lowermost ones (which, due to the periodic boundary conditions
coincide with the uppermost ones). We obtain, using eq.(\ref{c4}):
\eq
\exp(S_0) =  \prod_{\vec x,i}\left(1+\sum_{j={1\over 2}}^\infty
\left[\ibes{2j+1}\right]^{N_t}
~\chi_j\left(U_{\vec x;i}P_{\vec{x}+i}
U^\dagger_{\vec x;i}P^\dagger_{\vec x}\right)\right)~~,
\label{a1}
\en
where $P_{\vec x}$ is the open Polyakov line (whose trace is the
Polyakov loop) in the site $\vec x$ and $U_{\vec x;i}$ are the remaining
lowermost
spacelike links. Integrating also on $U_{\vec x;i}$
using eq.(\ref{c5})  we end up with
\eq
\label{a2}
\exp(S_0) = \prod_{\vec x,i}\left(1+\sum_{j={1\over 2}}^\infty
\left[\ibes{2j+1}\right]^{N_t}
~\chi_j(P_{\vec{x}+i}) ~\chi_j (P^\dagger_{\vec{x}})
\right) .
\en
Let us define, for future convenience, the link\footnote{The links we are
referring to are those of the $d$-dimensional spatial lattice, corresponding
to a space-like slice in the original $d+1$-dimensional
lattice} element of $\exp(S_0)$ as follows:
\eq
C^0_{\vec x; i} \equiv \sum_{j=0}^\infty
\left[\frac{I_{2j+1}(\beta_t)}{I_1(\beta_t)}\right]^{N_t}
~\chi_j (P_{\vec x+i})\chi_j (P^\dagger_{\vec{x}})~~.
\label{n2}
\en
It is now evident that this basic element, which will be denoted also as
$C^0_{\vec x, i} =
C^0(\theta_{\vec x},\theta_{\vec x+ i})$, depends only on
$\theta_{\vec x}, \theta_{\vec x +i}$, which are  the invariant angles
for the Polyakov lines
$P_{\vec x},P_{\vec x +i}$ in the sites joined by the link.
Indeed from now on we will always
assume to have gauge-rotated the Polyakov lines to be diagonal:
\eq
\label{zero2}
P_{\vec x}=\left(\begin{array}{cc} e^{\ii\theta_{\vec x}} & 0 \\
 0 & e^{-\ii\theta_{\vec x}}  \end{array}\right)
\en
Notice that the zero$^{\rm th}$-order action (\ref{a2})
is simply given by
\eq
\label{zero3}
\exp(S_0) = \prod_{\vec x ; i} C^0_{\vec x;i}
\en

\subsection{First order approximation}

The $O(\beta_s^0)$ effective action (\ref{zero3})
contains just nearest-neighbour
interactions between the Polyakov loops.
As we shall see in the following, the net outcome to the effective action
from the $O(\beta_s^2)$ terms in (\ref{gen2}) is the addition of
interaction terms involving more than two Polyakov loops.
Specifically, the term of order $\beta_s^2$ contains
interactions among the invariant angles of all the Polyakov lines
around a spatial plaquette.

Let us now compute the contributions from the adjoint
space-like plaquettes and from the pairs of fundamental ones in eq.
(\ref{gen2}).
As already said in the introduction, the next two subsections, containing 
these computations, are quite technical; the results are summarized in section
\ref{summary}.
\iffigs
\begin{figure}
\label{genfig1}
\epsfxsize 11cm
\begin{center}
\null\hskip 1pt\epsffile{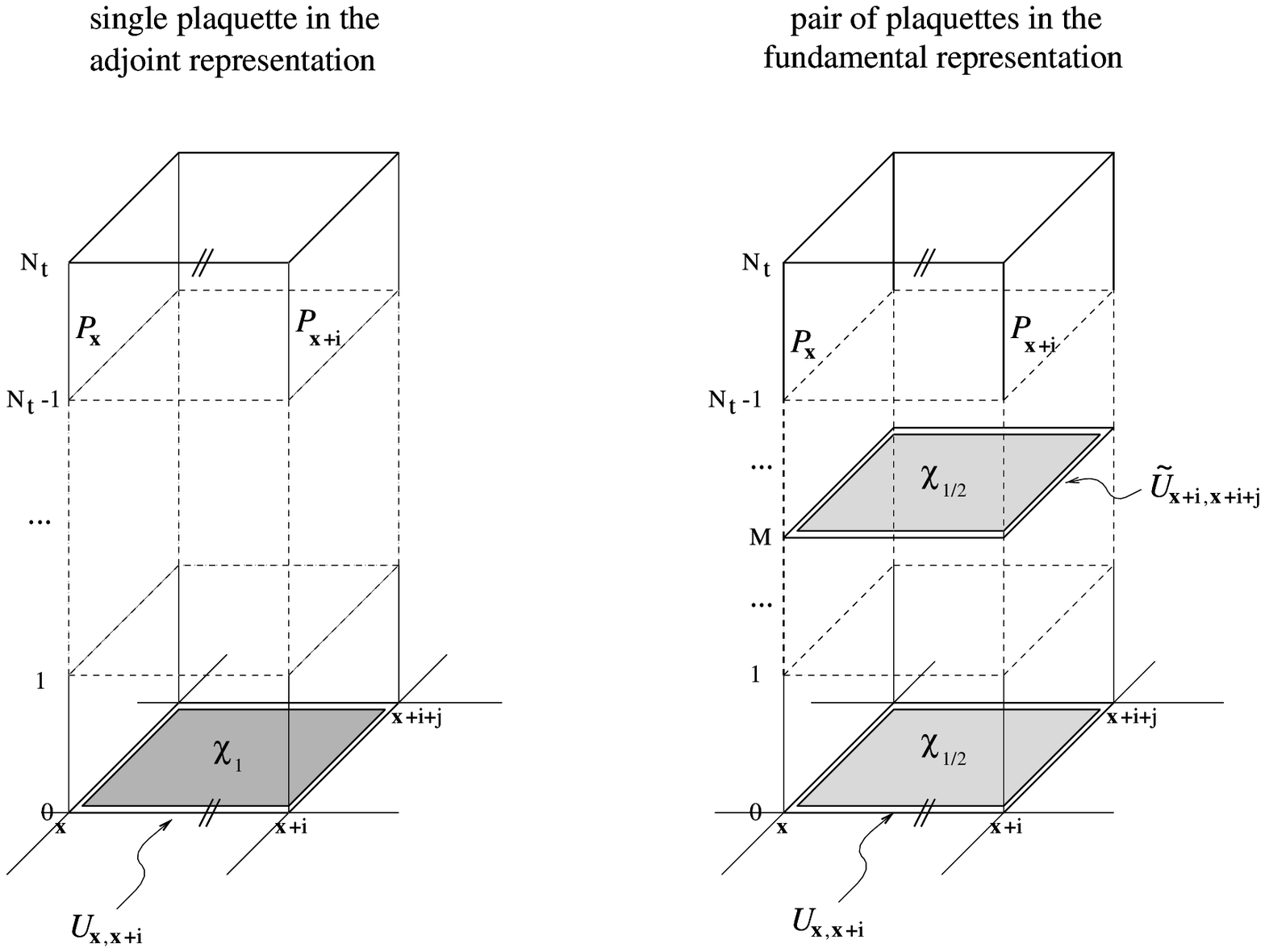}
\vskip 3pt\noindent
{\bf Fig. 1.} {\it Possible contributions to the effective action at
$O(\beta_s^2)$.}
\end{center}
\end{figure}
\fi

\subsubsection{The adjoint representation term}
\label{adjoint}
To calculate the contribution from the adjoint representation,
we have to select from  (\ref{gen2}) the term:
\eqa
&& 3 \ibessp{3}
\int \prod_{{\vec x}^\prime,t^\prime;i^\prime}
DU_{{\vec x}^\prime,t^\prime;i^\prime}\hskip 2pt
\prod_{{\vec x}^{\prime\prime}, t^{\prime\prime}; i^{\prime\prime}}
\left(1 + \sum_{j={1\over2}}^\infty d_j
\ibes{2j+1} \chi_j(U_{{\vec x}^{\prime\prime}, t^{\prime\prime};
0i^{\prime\prime}} ) \right) \nonumber \\
&&\times
\sum_{\vec x, i < j} \sum_{t=1}^{N_t} \chi_1(U_{\vec x, t; ij}) .
\label{adj1}
\ena
Let us fix a spatial position $\vec x,i<j$ in the above sum over the
space-like plaquettes, and study the corresponding integral. The
integration over all the link matrices not pertaining to the chosen
spatial position can be performed exactly as in the $O(\beta_s^0)$
case, giving as a result the product of all the 
link factors $C^0_{{\vec x}^\prime,i^\prime}$ except those in the chosen
spatial position\footnote{Later on we will denote this product
as $\prod_{{\rm link}\not\in {\rm pl}} C_0({\rm link})$.} 
(i.e. except $C^0_{{\vec x},i},C^0_{{\vec x +i},j},$
$C^0_{{\vec x+i+j},-i}, C^0_{{\vec x+j},-j}$).
To treat the remaining non-trivial integrations,
first we can note that all the spacelike
plaquettes in the same spatial position give evidently the same contribution,
regardless of the time $t$; therefore the sum over the time positions in
(\ref{adj1}) results simply in a $N_t$ factor. Secondly, it is convenient to
use the following relation for the $SU(2)$ characters:
\eq
\label{chi1}
\chi_1 = (\chi_{1\over 2})^2 - 1 .
\en
The ``$-1$'' simply reproduces the zeroth order
term, and gives a renormalization of order $\beta_s^2$ to such
contribution.
The integral over the link variables
along the plaquette can now be decoupled into products of
integrals over
single link matrices, by writing explicitly
$[\chi_{1 \over 2}(U_{\vec x,t ; ij})]^2$ 
in term of traces of link variables in the fundamental representation.
Thus eq.(\ref{adj1}) can be rewritten in terms of the following integrals over
the unitary spacelike link matrix $U$:
\eq
\label{beq}
B_{\alpha\beta\gamma\delta}(P_{\vec x},P_{\vec x+ i}) = \int DU \hskip 3pt
\left(1+\sum_{j={1\over 2}}^\infty d_j \left[\ibes{2j+1}\right]^{N_t}
\chi_j(U P_{\vec x +i} U^\dagger P_{\vec x}^\dagger )\right) U_{\alpha\beta}
U^\dagger_{\gamma\delta}
\en
where $\alpha,\ldots=1,2$ are the indices of the $U$ matrix in the
fundamental representation.
Let us assume that the Polyakov lines $P_{\vec x}$ in eq.(\ref{beq})
are already set in the diagonal form of eq.(\ref{zero2}).
Then the measure and  the argument of $\chi_j$ at the r.h.s.
of eq.(\ref{beq}) are invariant under the transformations
\begin{equation}
U_{\alpha\beta} \rightarrow \omega_{\alpha\alpha}~U_{\alpha\beta}~,
\phantom{ppppp}
U^\dagger_{\gamma\delta} \rightarrow
U^\dagger_{\gamma\delta}~ (\omega^{-1})_{\delta\delta}
\label{omegasym1}
\end{equation}
and
\begin{equation}
U_{\alpha\beta} \rightarrow U_{\alpha\beta} ~\omega_{\beta\beta}~,
\phantom{ppppp}
U^\dagger_{\gamma\delta} \rightarrow (\omega^{-1})_{\gamma\gamma}~
U^\dagger_{\gamma\delta}
\label{omegasym2}
\end{equation}
where $\omega$ is a  diagonal $SU(2)$ matrix. By using this invariance
one can easily conclude that
$B_{\alpha\beta\gamma\delta} = 0 $ unless $\beta = \gamma$ and $\delta =
\alpha$. As a consequence, the integral (\ref{beq}) depends on the invariant
angles of the Polyakov line only, and can be written as follows:
\eq
\label{beq2}
B_{\alpha\beta\gamma\delta}(P_{\vec x}, P_{\vec x + i})
\equiv B_{\alpha\beta\gamma\delta}(\theta_{\vec x}, \theta_{\vec x + i})
=\delta_{\beta\gamma}\delta_{\delta\alpha}
~C_{\alpha\beta}(\theta_{\vec x}, \theta_{\vec x + i})
\en
(no summation over repeated indices).
Moreover, it is not difficult to show that $C_{\alpha\beta}$ is a real
symmetric matrix.
By using these facts, we can write the contribution (\ref{adj1}) to the
effective action in terms of the invariant angles of the Polyakov lines.
The contribution to eq.(\ref{adj1})  at fixed $\vec x,i$ can be expressed
as:
\eq
3 N_t \frac{I_3(\beta_s)}{I_1(\beta_s)} ~
\biggl[ \prod_{\vec x^\prime, i^\prime} C^0_{\vec x^\prime,i^\prime} \biggr] ~
\biggl[
\trace [\widehat{C}(\theta_{\vec x},\theta_{\vec x +i})
\widehat{C}(\theta_{\vec x +i},\theta_{\vec x + i+j})
\widehat{C}(\theta_{\vec x +i+j},\theta_{\vec x + j})
\widehat{C}(\theta_{\vec x + j},\theta_{\vec x})] - 1
\biggr]
\label{adj2}
\en
where  the term (-1) in (\ref{adj2}) corresponds to the term (-1) in
(\ref{chi1}) and the matrices
$\widehat{C}(\theta_{\vec x},\theta_{\vec x + i})$
are a normalized version of $C$:
\eq
\label{chatadj}
\widehat{C}_{\alpha,\beta}(\theta_{\vec x},\theta_{\vec x+ i}) =
{C_{\alpha,\beta}(\theta_{\vec x},\theta_{\vec x+ i})
\over C^0_{\vec x; i}}
\en
Indeed in writing eq.(\ref{adj2}) we have multiplied and divided by
$C^0_{\vec x,i}\cdot$ $C^0_{\vec x+i,j}\cdot$ $C^0_{\vec x+i+j,-i}\cdot$
$C^0_{\vec x+j,-j}$ in order to collect the factor
$\prod_{\vec x, i} C^0_{\vec x,i}=\exp S_0$.

The last step is the explicit evaluation of the matrix elements
$C_{\alpha\beta}$.
We set
\eq
C_{\alpha\beta}=\sum_{j=0}^\infty
d_j~\left[\frac{I_{2j+1}(\beta_t)}{I_1(\beta_t)}\right]^{N_t}~
\cj_{\alpha\beta}~~,
\label{ab8bis}
\en
so that $\cj_{\alpha\beta}$ is defined as the contribution of the $j^{\rm th}$
representation in eq.(\ref{beq}):
\eq
\label{new1}
\int DU U_{\alpha\beta} U^\dagger_{\gamma\delta} \chi_j(UP_{\vx + i}U^\dagger
P^\dagger_\vx )
\equiv\delta_{\alpha\delta} \delta_{\gamma\beta} \cj_{\alpha\beta} .
\en
It follows from (\ref{new1}) that the non vanishing integrals at the
l.h.s. depend only on $|U_{\alpha\beta}|^2$, and hence that
$\cj_{11} = \cj_{22}$ and $\cj_{12} =\cj_{21}$.

To compute the matrix elements $\cj_{\alpha\beta}$, which is not a completely
trivial task, we use the following strategy\footnote{There is another
possible way to compute $\cj_{\alpha\beta}$, based on the expression
of the SU(2) characters as Tchebicheff polynomials of second kind:
$\chi_j(\theta)=U_{2j}\left(\cos(\theta)\right)$, which was utilized in 
\cite{var}. However this alternative
technique cannot easily be applied to the case of a pair of 
fundamental plaquettes.}.
We note that the matrix $\cj$ can be expressed in terms of the integral
\eq
\label{new2}
K^{(j)}(\theta_\vx,\theta_{\vx+i}) =
\int DU \chi_j(U P_2 U^\dagger P^\dagger_1)=
{1\over d_j} \chi_j(P_2)\chi_j(P^\dagger_1)
\en
through a system of two linear Schwinger--Dyson-like equations.
Indeed,
considering the integral $\int DU U_{\alpha\beta}$ $U^\dagger_{\beta\alpha}$
$\chi_j(UP_{\vx + i}U^\dagger P^\dagger_\vx)$ we easily find that
\eq
\label{newrel1}
\cj_{11} + \cj_{12}=K^{(j)} .
\en
To construct a second independent equation, let us consider the integral
\eq
\label{newint}
\int DU \chi_\um(UP_{\vx + i}U^\dagger P^\dagger_\vx)
\chi_j(UP_{\vx + i}U^\dagger P^\dagger_\vx).
\en
On one hand we can write the character $\chi_\um$ explicitly as a trace and
express the integral in terms of the $\cj_{\alpha\beta}$ by using
eq.(\ref{new1}). On the other hand, the integral (\ref{newint}) can be
written in terms of $K^{(j)}$ functions by using the basic SU(2)
Clebsch-Gordan relation: $\chi_{1\over 2} \chi_j$ $= \chi_{j+{1\over 2}} +
\chi_{j-{1\over 2}}$. The resulting equation is:
\eq
\label{newrel2}
2 \cos(\theta_{\vx + i} - \theta_\vx) \cj_{11} +
2\cos(\theta_{\vx + i} + \theta_\vx) \cj_{12} =
K^{(j-{1\over 2})} + K^{(j+{1\over 2})}
\en
Eq.s (\ref{newrel1}) and (\ref{newrel2}) form a set of two linear equations
in the two unknowns $\cj_{11}$ and $\cj_{12}$ whose solution is:
\eqa
\label{newsol}
\cj_{11} & = & {K^{(j-{1\over 2})} - 2 \cos(\theta_{\vx + i} + \theta_\vx)
K^{(j)} + K^{(j+{1\over 2})} \over 4 \sin \theta_{\vx + i} \sin \theta_\vx}
\nonumber\\
\cj_{12} & = & -{K^{(j-{1\over 2})} - 2 \cos(\theta_{\vx + i} - \theta_\vx)
K^{(j)} + K^{(j+{1\over 2})} \over 4 \sin \theta_{\vx + i} \sin \theta_\vx}
\ena
By inserting these results in eq.(\ref{ab8bis}) we finally
obtain the $C_{\alpha\beta}$ coefficients.
We choose to write the matrix $C$ in the form
\eq
C=\um\left(\matrix{C_0+C_1 & C_0 -C_1\cr C_0-C_1 & C_0 + C_1}\right)
\label{pec1}
\en
which is consistent with the symmetries of $C$ and 
allows an easy evaluation of the trace contained in eq.
(\ref{adj2}).
After some algebraic rearrangements it follows from eq.(\ref{newsol})
that
\eqa
C_1(\theta_\vx,\theta_{\vx + i}) & = &
{1\over {2\sin\theta_{\vx + i}\sin\theta_\vx}}
\biggl\{\sum_{j={1\over 2}}^\infty
\biggl[\chi_j(\theta_{\vx + i})\chi_j(\theta_\vx)
\frac{I_{2j+2}(\beta_t)^{N_t}-I_{2j}(\beta_t)^{N_t}}
{(2j+1)I_1(\beta_t)^{N_t}}\nonumber\\
& & +2\cos\left[(2j+1)\theta_{\vx + i}\right]\cos\left[(2j+1)\theta_\vx\right]
\left(\frac{I_{2j+1}(\beta_t)}{I_1(\beta_t)}\right)^{N_t}\biggr]
+\nonumber\\
& & +2 \cos\theta_{\vx + i} \cos\theta_\vx +
\left(\frac{I_2}{I_1}\right)^{N_t}\biggl\} ,
\label{ab19}
\ena
while $C_0(\theta_{\vx + i},\theta_\vx)$ coincides with the link element
$C^0_{\vx;i}$, as defined in eq.(\ref{n2}).

With the parametrization of the matrix $C$ given in eq.(\ref{pec1}) the 
trace in eq.(\ref{adj2}) simply reads: 
\eqa
\label{trace1}
&& \trace[C(\theta_\vx,\theta_{\vx + i})C(\theta_{\vx + i},\theta_{\vx +i+j})
C(\theta_{\vx +i+j},\theta_{\vx +j})C(\theta_{\vx +j},\theta_\vx)] \nonumber\\
&&
= C_0(\theta_\vx,\theta_{\vx +i})C_0(\theta_{\vx +i},\theta_{\vx +i+j})
C_0(\theta_{\vx +i+j},\theta_{\vx +j})C_0(\theta_{\vx +j},\theta_\vx)\nonumber\\
&& + C_1(\theta_\vx,\theta_{\vx +i})C_1(\theta_{\vx +i},\theta_{\vx +i+j})
C_1(\theta_{\vx +i+j},\theta_{\vx +j})C_1(\theta_{\vx +j},\theta_\vx).
\ena
By inserting this expression into eq.(\ref{adj2}), the products of $C_0$'s
cancel, and we are left with the explicit expression for the contribution of
an adjoint plaquette:
\begin{equation}
\label{resnew}
3 N_t \frac{I_3(\beta_s)}{I_1(\beta_s)} ~
\biggl[ \prod_{\vec x, i} C^0_{\vec x,i} \biggr]
{\hat C}_1(\theta_\vx,\theta_{\vx +i})
{\hat C}_1(\theta_{\vx +i},\theta_{\vx +i+j})
{\hat C}_1(\theta_{\vx +i+j},\theta_{\vx +j})
{\hat C}_1(\theta_{\vx +j},\theta_\vx) 
\end{equation}
where 
\eq
 {\hat C}_1(\theta_\vx,\theta_{\vx +i}) = {C_1(\theta_\vx,\theta_{\vx +i})
\over C^0_{\vec x,i}}
\en
and $C_1(\theta_\vx,\theta_{\vx +i})$ is given by eq.(\ref{ab19}). 

\subsubsection{Pair of fundamental representations}
\label{fundamental}
Let us go back to eq.(\ref{gen2}), and consider the last type of
contributions, namely the ones coming from two plaquettes in the
fundamental representation. As in
the previous case of the adjoint plaquettes, we consider the contribution
of a single pair of plaquettes; that is, we fix the spatial position $\vx,i<j$
of the plaquettes.
We can moreover fix the time position of one of these two plaquettes,
say at $t_1=0$.
Then the sum over $t_1$ just gives a factor of $N_t$.
The second spatial plaquette will be located at $t_2=M$, ($M=1,\ldots,
N_t-1$). Notice that the contribution from the
fundamental plaquettes is not present when $N_t=1$.

To perform the computation of this contribution, it is convenient to
choose the gauge so that the Polyakov lines are concentrated for instance
in the uppermost vertical links; this choice is always possible.

The integration of all the space-like links not involved in the two
space-like plaquettes can be performed in the usual way.
With notations similar to those of eq.(\ref{beq})
[see also Fig. 1]
we are left with the expression:
\eqa
\label{pair1}
&&
\prod_{{\rm link}\not\in{\rm pl}} C_0({\rm link})
\times 4 \Bigl(\ibessp{2}\Bigr)^2 N_t
\nonumber \\
&& \times
\sum_{M=1}^{N_t-1} \int \prod_{\vx\not =\vec y\in{\rm pl}} 
\biggl[ DU_{\vx,\vec y} D{\widetilde U}_{\vx,\vec y}
\hskip 2pt \biggl(1 + \sum_{m={1\over 2}}^\infty d_m
\left[\ibes{2m+1}\right]^M \chi_m(U_{\vx ,\vec y}
{\widetilde U}_{\vx ,\vec y}^\dagger)\biggr)
\nonumber\\
&& \times \biggl(1 + \sum_{n={1\over 2}}^\infty d_n
\left[\ibes{2l+1}\right]^{N_t - M} \chi_n(\widetilde{U}_{\vx ,\vec y}
P_{\vec y} U_{\vx, \vec y}^\dagger P_\vx^\dagger)\biggr)\biggr]\nonumber\\
&& \times
\trace (U_{\vx,\vx + i} U_{\vx + i,\vx +i+j} U_{\vx+j,\vx+i+j}^\dagger
U_{\vx,\vx+j}^\dagger)
\hskip 2pt \trace (\widetilde{U}_{\vx,\vx + i}\widetilde{U}_{\vx + i,\vx +i+j}
\widetilde{U}_{\vx+j,\vx+i+j}^\dagger \widetilde{U}_{\vx,\vx +j}^\dagger) .
\nonumber\\
\ena
By writing explicitly the products in the traces,
one can express eq.(\ref{pair1}) in terms of integrals of the form
\eq
\label{fun4}
\int DU\hskip 2pt D\widetilde U \hskip 3pt U_{\alpha\beta}
{\widetilde U}^\dagger_{\gamma\delta}
\chi_m(U{\widetilde U}^\dagger)\chi_n({\widetilde U} P_{\vx +i} U^\dagger
P_\vx^\dagger)
\equiv \delta_{\alpha\delta} \delta_{\beta\gamma}\hskip 3pt
{\cal C}^{(m,n)}_{\alpha\beta}(\theta_\vx,\theta_{\vx +i}) .
\en
where the Kronecker deltas at the r.h.s. originate, as in the case of
the adjoint representation, by the symmetry given in eq.s
(\ref{omegasym1},\ref{omegasym2}). We have assumed in eq.(\ref{fun4})
to have diagonalized the Polyakov lines as in eq.(\ref{zero2}).
The original expression (\ref{pair1}) can then be written as
\eqa
\label{pair2}
&& \prod_{{\rm link}\not\in{\rm pl}} C_0({\rm link})\times
4 N_t \Bigl(\ibessp{2}\Bigr)^2\nonumber \\
&& \times \sum_{M} \trace [
C^{(M)}(\theta_\vx,\theta_{\vx +i})
C^{(M)}(\theta_{\vx +i},\theta_{\vx +i+j})
C^{(M)}(\theta_{\vx +i+j},\theta_{\vx +j})
C^{(M)}(\theta_{\vx +j},\theta_\vx)].
\nonumber\\
\ena
where the $2\times 2$ matrices $C^{(M)}$ are given by
\eq
\label{fun3}
C^{(M)}_{\alpha\beta}= \sum_{m,n} d_m d_n \left[\ibes{2m+1}\right]^M
\left[\ibes{2n+1}\right]^{N_t-M} {\cal C}^{(m,n)}_{\alpha\beta}.
\en
Unlike the ${\cal C}^{(j)}_{\alpha\beta}$ defined in eq.(\ref{new1}),
${\cal C}^{(m,n)}_{\alpha\beta}$ is not a real symmetric
matrix.
It rather satisfies the following properties:
\eq
\label{fun4a}
\left[{\cal C}^{(m,n)}_{11}\right]^* = {\cal C}^{(m,n)}_{22}
\hskip 0.5cm ; \hskip 0.5cm
\left[{\cal C}^{(m,n)}_{12}\right]^* = {\cal C}^{(m,n)}_{21} .
\en
Moreover, one can easily show from its definition (\ref{fun4}) that
\eq
\left[{\cal C}^{(m,n)}_{\alpha\beta}\right]^* =
(P_{\vx +i})_{\beta\beta} (P_\vx^\dagger)_{\alpha\alpha}
{\cal C}^{(n,m)}_{\alpha\beta}
\label{cnew}
\en
and ${\cal C}^{(m,n)}_{\alpha\beta}(\theta_\vx,\theta_{\vx +i}) =
{\cal C}^{(m,n)}_{\beta\alpha}(-\theta_\vx ,-\theta_{\vx +i})$.

In order to compute the matrix elements of ${\cal C}^{(m,n)}$
we follow a strategy
analogous to that utilized in the adjoint plaquette case.
We show how the ${\cal C}^{(m,n)}_{\alpha\beta}$, that have four independent
real components, can be expressed in term of the integral
\eq
\label{fun5}
\int DU\hskip 2pt D\widetilde U \hskip 3pt
\chi_m(U{\widetilde U}^\dagger) \chi_n({\widetilde U}
P_{\vx +i} U^\dagger P_\vx^\dagger) = {\delta_{mn}\over d_n^2} \chi_n
(\theta_{\vx +i})\chi_n (\theta_\vx)\equiv {\delta_{mn}\over d_n}K^{(n)} 
\en
by means of a set of four linear Schwinger--Dyson-like equations.
Consider first the integral
\eq
\int DU\hskip 2pt D\widetilde U \chi_{1\over
2}(U{\widetilde U}^\dagger) \chi_m(U{\widetilde U}^\dagger)
\chi_n(\widetilde U P_{\vx +i} U^\dagger P_\vx^\dagger) .
\label{intnew}
\en
On one hand we can write explicitly
as a trace the $\chi_{1\over 2}$ factor and use the definition
(\ref{fun4}); on the other we can use the Clebsch-Gordan relation to rewrite
the integral (\ref{intnew}) as a combination of $K^{(n)}$ functions
via eq.(\ref{fun5}).
We thus find:
\eq
\label{uno}
\cmn 11 + \cmn 12 + \cmn 22 + \cmn 21 = (\delta_{m+{1\over 2},n} +
\delta_{m-{1\over 2},n}) {1\over d_n} K^{(n)}.
\en
By considering an integral analogous to (\ref{intnew}), but containing the
trace $\chi_{1\over 2} (\widetilde U P_{\vx +i} U^\dagger P_\vx^\dagger)$
instead of $\chi_{1\over 2}(U{\widetilde U}^\dagger)$, we obtain that
\eqa
\label{due}
&&\ee{\ii(\theta_{\vx +i}-\theta_\vx)} \cmn 11 +
\ee{-\ii (\theta_{\vx +i} - \theta_\vx)}
\cmn 22 + \ee{\ii(\theta_\vx + \theta_{\vx +i})} \cmn 21 +
\ee{-\ii(\theta_\vx +
\theta_{\vx +i})} \cmn 12 = \nonumber\\
&&\hskip 1cm = (\delta_{m,n+\um} + \delta_{m, n -\um}){1\over d_m}
K^{(m)} .
\ena
To determine all the components of $\cmn \alpha\beta$ we need two more
relations, that can be obtained by taking derivatives
of eq.(\ref{fun5}) with respect to the invariant angles\footnote{SU(2)
characters satisfy
${\partial\over\ \partial \chi_\um(\theta)}$
$\bigl[ \chi_{j +\um}(\theta) - \chi_{j -\um}
(\theta)\bigr]$ $ = (2j + 1) \chi_j(\theta)$,
as can be seen by expressing them as Tchebicheff polynomials:
$\chi_j(\theta) \equiv U_{2j}(\cos(\theta))$.
One has then for instance:
\begin{displaymath}
{\partial\over\partial\theta_{\vx +i}}\bigl[\chi_{n+\um} -\chi_{n-\um}\bigr]
(\widetilde U P_{\vx +i} U^\dagger P_\vx^\dagger) =
(2n+1)\chi_n(\widetilde U P_{\vx +i} U^\dagger P_\vx^\dagger) {\partial\over
\partial\theta_{\vx +i}} \trace
(\widetilde U P_{\vx +i}U^\dagger P_\vx^\dagger).
\nonumber\\
\end{displaymath}
}:
\eq
\label{tre}
\ii(2n+1)\bigl[\ee{\ii(\theta_\vx+\theta_{\vx +i})}\cmn 21 -
\ee{-\ii(\theta_\vx+\theta_{\vx +i})}\cmn 12 \bigr]
= {1\over 2d_m}
(\delta_{m,n+\um} - \delta_{m,n-\um}) \partial_+
K^{(m)}
\en
and
\eq
\label{quattro}
\ii(2n+1)\bigl[\ee{\ii(\theta_{\vx +i}-\theta_\vx)}\cmn 11 -
\ee{-\ii(\theta_{\vx +i}-\theta_\vx)}\cmn 22 \bigr]
= {1\over 2d_m}
(\delta_{m,n+\um} - \delta_{m,n-\um}) \partial_-
K^{(m)},
\en
where $\partial_\pm$ stands for ${\partial \over\partial\theta_{\vx +i}}
\pm {\partial \over\partial\theta_\vx} $.
We can now obtain the expression of $\cmn \alpha\beta$ by solving the system
formed by the four equations (\ref{uno},\ref{due},\ref{tre},\ref{quattro}).
This is more easily done in terms of the matrix
${\widetilde{\cal C}}^{(m,n)}$, defined by
$\ctmn_{11} = \ee{\ii(\theta_{\vx +i} -\theta_\vx)} \cmn 11$ and $\ctmn_{12} =
\ee{-\ii(\theta_\vx + \theta_{\vx +i})}\cmn 12$, i.e. by
\eq
\label{defctilde}
\ctmn_{\alpha\beta} = (P_{\vx +i})_{\beta\beta} (P_\vx)_{\alpha\alpha}
\cmn {\alpha}{\beta} .
\en
The matrix $\ctmn_{\alpha\beta}$ enjoys the same symmetries (\ref{fun4a})
as $\cmn {\alpha}{\beta}$, while eq.(\ref{cnew}) is replaced by
\eq
\label{ctildenew}
\left[\ctmn_{\alpha\beta}\right]^* = (P_{\vx + i}^\dagger)_{\beta\beta}
(P_\vx)_{\alpha\alpha} {\widetilde {\cal C}}^{(n,m)}_{\alpha\beta} .
\en
The symmetries (\ref{fun4a}) can be implemented by writing the 
$2\times 2$ matrix ${\widetilde C}^{(m,n)}$ as
\eq
\label{cmnform}
\ctmn = \um \left(\matrix{\czmn +\cumn -\ii \bmmn & \czmn - \cumn + \ii
\bpmn \cr
\czmn - \cumn  -\ii \bpmn & \czmn +\cumn +\ii \bmmn}\right) .
\en
The equations (\ref{uno},\ref{due},\ref{tre},\ref{quattro}) imply that the
only non-vanishing components of $\ctmn$ are those with $n = m \pm \um$.
The solution of the system can be expressed as follows:
\eqa
\label{sol1}
&&\cz m{m+\um} = \cz m{m-\um} =
{1\over 2 d_m} K^{(m)}
\nonumber\\
&& d_m \bpm m{m+\um} = -{1\over 2(2m+2)} \partial_\pm K^{(m)}
\hskip 0.3cm ; \hskip 0.3cm
d_m\bpm m{m-\um} = {1\over 2(2m)} \partial_\pm K^{(m)}
\nonumber\\
&& d_m \cu m{m+\um} =
{\cos\left[(2m+1)\theta_{\vx +i}\right]
\cos\left[(2m+1)\theta_\vx\right]
+\cos \theta_{\vx+i} \cos\theta_{\vx} K^{(m)}
- K^{(m+\um)} \over
(2m+2)\hskip 3pt 2\sin\theta_{\vx +i} \sin\theta_\vx} \nonumber\\
&&d_m \cu m{m-\um} =
{\cos\left[(2m+1)\theta_{\vx +i}\right]
\cos\left[(2m+1)\theta_\vx\right]
-\cos \theta_{\vx+i} \cos\theta_\vx K^{(m)}
+ K^{(m-\um)}\over
(2m) \hskip 3pt2\sin\theta_{\vx +i} \sin\theta_\vx}
\nonumber\\
\ena

In analogy to eq.(\ref{fun3}) we can introduce a matrix
${\widetilde C}^{(M)}_{\alpha\beta}$ defined by the relation

\eqa
\label{fun33}
{\widetilde C}^{(M)}_{\alpha\beta}(\theta_{\vx +i},\theta_\vx) & = &
(P_{\vx +i})_{\beta\beta} (P_\vx^\dagger)_{\alpha\alpha}
C^{(M)}(\theta_{\vx +i},\theta_\vx)\nonumber \\ 
& = & \sum_{m,n} d_m d_n \left[\ibes{2m+1}\right]^M
\left[\ibes{2n+1}\right]^{N_t-M} {\widetilde{\cal C}}^{(m,n)}_{\alpha\beta}
(\theta_{\vx +i},\theta_\vx).
\ena 

We can now insert the matrix ${\widetilde C}^{(M)}$ instead of
$C^{(M)}$ in the expression (\ref{pair2}), as the
extra phases appearing in ${\widetilde C}^{(M)}$ cancel in the trace. 

The matrix ${\widetilde C}^{(M)}_{\alpha\beta}$ can be written, in analogy with eq.
(\ref{cmnform}), as
\begin{equation}
\label{newc}
{\widetilde C}^{(M)} = \um \left(\matrix{C^{(M)}_0 + C^{(M)}_1 -\ii B^{(M)}_-
& C^{(M)}_0 - C^{(M)}_1 +\ii B^{(M)}_+ \cr
C^{(M)}_0 - C^{(M)}_1 -\ii B^{(M)}_+ &
C^{(M)}_0 + C^{(M)}_1 +\ii B^{(M)}_-}\right) .
\end{equation}
The explicit form of $C^{(M)}_0,C^{(M)}_1$ and $B^{(M)}_\pm$ can
now be obtained from eq.s (\ref{fun33}) and (\ref{sol1}).
The result is:
\eqa
\label{sol2}
C^{(M)}_0 & = & \um\sum_{m=\um}^{\infty}
 \ibe{2m+1}M \left(d_{m+\um}\ibe{2m+2}{N_t-M}  +
d_{m-\um}\ibe{2m}{N_t-M}\right)
K^{(m)}(\theta_{\vx +i},\theta_\vx)\nonumber\\
&&+\ibe 2{N_t-M}\nonumber\\
B_\pm^{(M)} & = & -\um\sum_{m=\um}^\infty
 \ibe{2m+1}M \left(\ibe{2m+2}{N_t-M} -
\ibe{2m}{N_t-M}\right) \partial_\pm
K^{(m)}(\theta_{\vx +i},\theta_\vx)\nonumber\\
C^{(M)}_1 & = & {1\over 2\sin\theta_{\vx +i}\sin\theta_\vx}\Biggl\{
\ibe 2M + \ibe 2{N_t-M} 2 \cos\theta_{\vx+i}\cos\theta_\vx
\nonumber\\
&& + \sum_{m=\um}^\infty \Biggl(\ibe {2m+1}{N_t-M} \left( \ibe{2m+2}M -
\ibe{2m}M\right)  K^{(m)}(\theta_{\vx +i},\theta_\vx) + \nonumber\\
&& +  \ibe {2m+1}{M} \left( \ibe{2m+2}{N_t-M} +
\ibe{2m}{N_t-M}\right) \Bigl(\cos\left[(2m+1)\theta_{\vx +i}\right]
\cos\left[(2m+1)\theta_\vx\right]\Bigr)
\nonumber \\
&& +  \ibe {2m+1}{M} \left( \ibe{2m+2}{N_t-M} -
\ibe{2m}{N_t-M}\right) \Bigl(
\cos \theta_{\vx+i}\cos\theta_\vx \hskip 3pt K^{(m)}(\theta_{\vx+i},\theta_\vx)
\Bigr)\Biggr)\Biggr\}.\nonumber\\
\ena
In the above formula, the argument of all the Bessel functions is
the timelike-coupling $\beta_t$.

The explicit expression (\ref{sol2}) 
of the matrix ${\widetilde C}^{(M)}$ is quite complicated; but there
are some non-trivial consistency checks that we can perform on it.
First, it is not difficult to see that if we set $M=0$ in the formulae
(\ref{sol2}), namely if we let the two fundamental plaquettes
coincide at the same time position, we correctly reproduce the
results obtained in section \ref{adjoint} for the $\chi_\um\chi_\um$
term in the adjoint plaquette:
\begin{eqnarray}
\label{mzero}
C^{(M)}_0(\theta_{\vec x},\theta_{\vec x+i}) & 
\stackrel{M \rightarrow 0}{\longrightarrow} &
C_0(\theta_{\vec x},\theta_{\vec x+i})\nonumber\\
C^{(M)}_1(\theta_{\vec x},\theta_{\vec x+i}) & 
\stackrel{M \rightarrow 0}{\longrightarrow} &
C_1(\theta_{\vec x},\theta_{\vec x+i})\nonumber\\
B_\pm^{(M)}(\theta_{\vec x},\theta_{\vec x+i}) & 
\stackrel{M \rightarrow 0}{\longrightarrow} &
0 .
\end{eqnarray}
Second, we can rearrange, through some algebraic manipulations, the
expressions (\ref{sol2}) in such a way that the symmetry under the
exchange of $M$ with $N_t-M$ :
\begin{equation}
\label{cmsym}
\bigl[{\widetilde C}^{(M)}_{\alpha\beta}(\theta_{\vec x},
\theta_{\vec x+i})\bigr]^* =
(P_{\vec x +i}^\dagger)_{\beta\beta} (P_{\vec x})_{\alpha\alpha}
{\widetilde C}^{(N_t-M)}_{\alpha\beta}(\theta_{\vec x},\theta_{\vec x+i}) ,
\end{equation}
which comes from the analogous property (\ref{ctildenew}) of 
$\ctmn_{\alpha\beta}$,
becomes manifest. It is possible indeed to write ${\widetilde C}^{(M)}$ 
as a matrix in the following form:
\begin{eqnarray}
\label{altc1}
{\widetilde C}^{(M)} & = &
{1\over 4\sin\theta_\vx\sin\theta_{\vx +i} }
\Biggl\{
\sum_{m=\um}^{\infty}
\Biggl\{
\left(\matrix {\ee{-\ii(2m+1)\theta_-} & -\ee{\ii(2m+1)\theta_+} \cr
-\ee{-\ii(2m+1)\theta_+} & \ee{\ii(2m+1)\theta_-} }\right)
\ibe{2m+1}M \ibe{2m+2}{N_t-M} \nonumber\\
&&  
+\left(\matrix {\ee{\ii(2m+1)\theta_-} & -\ee{-\ii(2m+1)\theta_+} \cr
-\ee{\ii(2m+1)\theta_+} & \ee{-\ii(2m+1)\theta_-} }\right)
\ibe{2m+1}M \ibe{2m}{N_t-M} \nonumber\\
&&+ K^{(m)}\pmatrix{1 & -1 \cr -1 & 1} \ibe{2m+1}{N_t-M} 
\Biggl(\ibe{2m+2}M -\ibe{2m}M\Biggl)\nonumber\\
&&+ K^{(m)}\pmatrix{\ee{\ii\theta_-} & -\ee{-\ii\theta_+} \cr 
-\ee{\ii\theta_+} & \ee{-\ii\theta_-} } \ibe{2m+1}{M} 
\Biggl(\ibe{2m+2}{N_t-M} -\ibe{2m}{N_t-M}\Biggl)\Biggl\}
\nonumber\\
&& +  \pmatrix{1 & -1\cr -1 & 1} \ibe{2}{M}
+  \pmatrix{\cos\theta_- & -\cos\theta_+ \cr 
-\cos\theta_+ & \cos\theta_-} \ibe 2{N_t-M} 
\Biggl\}
\end{eqnarray}

Finally, having determined the matrix 
${\widetilde C}^{(M)}(\theta_{\vx+i},\theta_\vx)$
we can insert its expression into eq.(\ref{pair2}), as we discussed above,
to get the contribution of a pair of fundamental plaquettes to the 
effective action.
The result of the trace is cumbersome and not very illuminating, so
we shall not write it here; it will however be used in the mean field
analysis of the following sections.

\subsection{The effective action up to $O(\beta_s^2)$}
\label{summary}
Let us summarize here our results by reporting the form of the
effective action for the Polyakov loops determined in the previous sections.
To this action we will in the later sections
apply standard and improved mean field techniques in order to extract the
value of the critical coupling. We have:
\begin{equation}
\exp S_{\rm eff} = \exp (S_0+S_1)
\label{act1}
\end{equation}
where
\begin{equation}
\label{act2}
\exp S_0=\sum_{x,i} C_0(\theta_{\vec x},\theta_{\vec x+i})
\end{equation}
and
\begin{eqnarray}
\label{act3}
S_1 &=& N_t \sum_{\vec x,i<j} \Biggl\{3  \ibessp{3}\hskip 8pt
{C_1(\theta_{\vec x},\theta_{\vec x+i}) C_1(\theta_{\vec x +i},
\theta_{\vec x+i+j}) C_1(\theta_{\vec x +i+j}, \theta_{\vec x+j})
C_1(\theta_{\vec x +j}, \theta_{\vec x}) \over
C_0(\theta_{\vec x},\theta_{\vec x+i}) C_0(\theta_{\vec x +i},
\theta_{\vec x+i+j}) C_0(\theta_{\vec x +i+j}, \theta_{\vec x+j})
C_0(\theta_{\vec x +j}, \theta_{\vec x})}\nonumber\\
&&+4 \Bigl(\ibessp{2}\Bigr)^2 \sum_{M=1}^{N_t-1}
{\trace \Bigl[{\widetilde C}^{(M)}(\theta_{\vec x},\theta_{\vec x+i})\ldots
{\widetilde C}^{(M)}(\theta_{\vec x+j},\theta_{\vec x})\Bigr]\over
C_0(\theta_{\vec x},\theta_{\vec x+i})\ldots
C_0(\theta_{\vec x+j},\theta_{\vec x})}\Biggr\}.
\end{eqnarray}
The quantities in the above equations are defined as follows:
$C_0(\theta_{\vec x},\theta_{\vec x+i})\equiv C^0_{\vec x,i}$ is
given by eq.(\ref{n2}), $C_1(\theta_{\vec x},\theta_{\vec x+i})$ 
by eq.(\ref{ab19}) and the matrix
${\widetilde C}^{(M)}(\theta_{\vec x},\theta_{\vec x+i})$  by equations
(\ref{newc}) and (\ref{sol2}) or, alternatively, by eq.(\ref{altc1}).

$S_0$ is the $O(\beta_s^0)$ effective action. 
It contains a sum over the
links $\vec x, i$ in the $d$-dimensional spatial lattice and each term of the sum
represents a nearest
neighbour interaction between Polyakov loops. $S_1$ describes
the effect at $O(\beta_s^2)$ of the space-like plaquettes, and it is
given by a sum over the plaquettes $\vec x, i<j$ in the
$d$-dimensional lattice. Each term represents the interaction among the four 
Polyakov loops at the vertices of each plaquette. Notice that in 
eqs. (\ref{act1}-\ref{act3}) the contribution of the space-like plaquettes 
has been exponentiated, which is correct at the order $O(\beta_s^2)$ .

\subsection{The $N_t=1$ case and the Kazakov-Migdal model}
The interesting feature of the $N_t=1$ case is that the
model that we are studying becomes a particular case of the
Kazakov-Migdal model \cite{kazmig}. This connection was already noticed
in~\cite{cdpht,cdmp2d}
and was the origin of our previous analysis in the $N\to\infty$
limit~\cite{bcdmp}.
All the integrals that we have described in the
previous sections can be directly evaluated in the $N_t = 1$ case as particular
instances of
a  nontrivial generalization of the so called Itzykson-Zuber integral
\cite{izhc},
evaluated in~\cite{kmsw}.
This alternative derivation in the $N_t=1$ case provides another 
check of our computations (at least for the 
contribution of the adjoint plaquettes, as this is the only 
$O(\beta_s^2)$ contribution when $N_t=1$).

The basic integral used in the  computation of $S_0$
[see eq.(\ref{zero1})] coincides, when $N_t=1$, with
the link integral
\begin{eqnarray}
\int d U_{\vec{x};i} \exp \left\{ {\beta_t \over 2}
{\rm Tr}_f \left( V(\vec{x}) U_{\vec{x};i} V^\dagger(\vec{x}+i)
U^\dagger_{\vec{x};i} \right) \right\} & = &
\nonumber \\
= { e^{\beta_t \cos(\theta_{\vec{x}} -\theta_{\vec{x}+i})}
- e^{ \beta_t \cos(\theta_{\vec{x}} +\theta_{\vec{x}+i})}
\over 2 \beta_t \sin(\theta_{\vec{x}}) \sin(\theta_{\vec{x}+i}) }
& \phantom{=}~, &
\label{a6}
\end{eqnarray}
which was non-perturbatively
computed in \cite{kmsw}. We can compare this expression with our general
result (\ref{a2}). We find that, if $N_t=1$, the character expansion 
contained in eq.(\ref{a2}) can be summed exactly. 
In fact by inserting  the explicit form (\ref{c1}) of 
the characters into eq.(\ref{a2}) and then using the relation
\eq
2~\sin[(2r+1)\theta_{\vec x}]~\sin[(2r+1)\theta_{\vec x+i}]=
\cos[(2r+1)(\theta_{\vec x}-\theta_{\vec x+i})]-
\cos[(2r+1)(\theta_{\vec x}+\theta_{\vec x+i})]
\label{a2ter}
\en
and the well known expansion
\eq
e^{\beta\cos{\theta}}=I_0(\beta)+2\sum_{k=1}^\infty
I_k(\beta)\cos(k\theta)~~~,
\label{a2quat}
\en
it is easy to obtain:
\eq
\exp(S_0) =  \prod_{\vec x,i}
\frac{e^{\beta_t \cos(\theta_{\vec x}-\theta_{\vec x+i})}-
e^{\beta_t \cos(\theta_{\vec x}+\theta_{\vec x+i})}}
{4I_1(\beta_t) \sin(\theta_{\vec x}) \sin(\theta_{\vec x+i})}~~.
\label{a3}
\en
This expression coincides with eq.(\ref{a6}), except for the irrelevant 
overall factor ${2 I_1(\beta_t)\over\beta_t}$ [see the remark after 
eq.(\ref{wilsonexp})]. 

In the $N_t=1$ case, the first non-trivial contributions from the 
space-like plaquettes, i.e. those
coming from an adjoint plaquette, can be extracted from the definition 
of the correlators
given in~\cite{kmsw} :
\begin{eqnarray}
& &
\left<
\left( U_{\vec{x};i} \right)_{\mu,\nu} \left( U^\dagger_{\vec{x};i}
\right)^{\rho,\sigma} \right>
 \nonumber \\
&\equiv  &
\frac{ \int d U_{\vec{x};i} \exp \left\{
{\beta_t \over 2} {\rm Tr}_f \left( P_{\vec{x}} U_{\vec{x};i}
PV^\dagger_{\vec{x}+i}
U^\dagger_{\vec{x};i} \right) \right\}
\left( U_{\vec{x};i} \right)_{\mu,\nu} \left( U^\dagger_{\vec{x};i}
\right)^{\rho,\sigma} }
{\int d U_{\vec{x};i} \exp \left\{
{\beta_t \over 2} {\rm Tr}_f \left( P_{\vec{x}} U_{\vec{x};i}
P^\dagger_{\vec{x}+i} U^\dagger_{\vec{x};i} \right) \right\} } .
\label{a7}
\end{eqnarray}
These correlators correspond to the $B_{\mu\nu\rho\sigma}$ of eq.(\ref{beq}),
divided by $C^0_{\vx,i}$; they must be therefore  diagonal,
namely:
\eq
\left<
\left( U_{\vec{x};i} \right)_{\mu,\nu} \left( U^\dagger_{\vec{x};i}
\right)^{\rho,\sigma} \right>
 =  \delta_\mu^\sigma \delta_\nu^\rho \hat C_{\mu,\nu} (\vec{x};i)
\en
where the $\hat C_{\mu,\nu}(\vec{x};i)$ are equivalent, apart from the
different normalization, to our $C_{kl}$ matrix elements. 
These correlators  were calculated in ~\cite{kmsw}:
\begin{eqnarray}
\hat C_{1,1}(\vec{x};i) & = & \hat C_{2,2}(\vec{x};i) =
{ 2\beta_t \sin(\theta_{\vec{x}}) \sin(\theta_{\vec{x}+i}) -
 \left( 1 - e^{-2\beta_t \sin(\theta_{\vec{x}}) \sin(\theta_{\vec{x}+i})}
\right) \over
\left( 1 - e^{-2\beta_t \sin(\theta_{\vec{x}}) \sin(\theta_{\vec{x}+i})}
\right) \left( 2\beta_t \sin(\theta_{\vec{x}}) \sin(\theta_{\vec{x}+i})
\right) }
\nonumber \\
\hat C_{1,2}(\vec{x};i) & = &\hat C_{2,1}(\vec{x};i)  =
{ 
1 - e^{-2\beta_t \sin(\theta_{\vec{x}}) \sin(\theta_{\vec{x}+i})}
\left(1+ 2\beta_t \sin(\theta_{\vec{x}}) \sin(\theta_{\vec{x}+i})
\right)
\over
\left( 1 - e^{-2\beta_t \sin(\theta_{\vec{x}}) \sin(\theta_{\vec{x}+i})}
\right) \left( 2\beta_t \sin(\theta_{\vec{x}}) \sin(\theta_{\vec{x}+i})
\right) }
\label{a8}
\end{eqnarray}
Again we can compare this result with our results for generic $N_t$, 
expressed in section \ref{adjoint} in terms of character expansions.
In the $N_t=1$ case
the sum over the representations can be performed exactly, and a closed
expression for the $C_{\alpha\beta}$ coefficients can be obtained.
This can be done by using the identity:
\eq
\label{idbes1}
I(\beta)_{n-1}-I(\beta)_{n+1}=2nI(\beta)_{n}
\en
and eq.(\ref{a2quat}).
The result is:
\eqa
C_{11}(N_t=1) & = &
\frac{e^{\beta_t \cos(\theta_\vx -\theta_{\vx + i})}}
{4I_1(\beta_t) \sin(\theta_\vx ) \sin(\theta_{\vx + i})}
-\frac{e^{\beta_t \cos(\theta_\vx -\theta_{\vx + i})}-
e^{\beta_t \cos(\theta_\vx +\theta_{\vx + i})}}
{8\beta_t I_1(\beta_t) \sin^2(\theta_\vx)
\sin^2(\theta_{\vx + i})}\nonumber\\
C_{12}(N_t=1) & = &
\frac{e^{\beta_t \cos(\theta_\vx -\theta_{\vx + i} )}-
e^{\beta_t \cos(\theta_\vx  +\theta_{\vx + i})}}
{8\beta_t I_1(\beta_t) \sin^2(\theta_\vx)
\sin^2(\theta_{\vx + i})}-
\frac{e^{\beta_t \cos(\theta_\vx+\theta_{\vx + i})}}
{4I_1(\beta_t) \sin(\theta_\vx ) \sin(\theta_{\vx + i})},
\label{an2}
\ena
which perfectly matches eq.(\ref{a8}), if we take into account the
normalization by $C^0_{\vx ,i}$ and the ${2 I_1(\beta_t)\over \beta_t}$
factor, as in eq.(\ref{a3})

\section{Mean Field computation of the critical coupling}
The effective action obtained in the previous section describes a $d$
dimensional  spin model with complicated interactions and
cannot be solved exactly. However several features of the model can be figured
out rather easily. First, it can be seen that all the interaction terms
are even functions of the variables $\theta$, so that the model has a global
$Z_2$ symmetry, in agreement with the Svetitsky-Yaffe conjecture. 
Here and in the following
we shall assume to have fixed the asymmetry ratio $\rho$, and 
we shall study the phase diagram
of the model in terms of only one coupling (for instance: $\beta_t$).
However, as already anticipated,
 the $\rho$ dependence of the critical temperature will
play  a major role.
 For large values of the
coupling $\beta_t$ the $Z_2$
 symmetry is spontaneously broken, hence we expect a
phase transition for some critical value of the coupling. The simplest way to
estimate this critical coupling is certainly the mean field approximation. As it
is well known this method gives in general rather rough estimates of the
critical temperature, and much more refined techniques have been elaborated in
these last years. In the section 4.4
we shall comment on this point in more detail and  apply an
improved version of the mean field approximation (Bethe approximation) to the
$N_t=1$ case. This test will make us confident of the the fact that the errors
which we make by keeping a plain mean field approximation are of the order of the
10\%-15\%. Nevertheless, in this section we shall restrict ourselves to the
 plain mean field approximation. The reason is twofold: first, a relevant part
 of our results (and in particular the agreement with the weak coupling
 calculation of Karsch) rely on {\sl differences of critical couplings}, and
 these differences are only slightly affected by the mean field approximation,
 which essentially affects the data in the sense of giving
 an overall systematic error. Second, as $N_t$
 increases
 the error that we make by using the mean field approximation
 becomes smaller than that
due to the truncation of the $\beta_s$ expansion. The huge amount of complexity
needed to implement more refined approximations would be justified, and
would become meaningful, only if higher orders in the expansion were  taken
into account. This could well be done in principle but, as we shall see, the
results  we obtain by keeping only the mean field result are already
very interesting, since they clearly show the expected trend.
Let us also mention that there is another situation in which more precise
methods to estimate the critical coupling are justified, namely the $N_t=1$
case, in which the diagrammatic
entropy of higher order contributions is highly constrained and we can expect
that the $\beta^2$ order alone gives already a very good approximation. This
will be the subject of sect.~4.4~~.

\subsection{Theoretical expectations.}
The ultimate test of the correctness of any lattice regularization
is that, as the continuum limit is approached, the various dimensional
quantities in which one is interested follow the correct scaling behaviour.
This scaling behaviour can be easily obtained by writing explicitly the
dependence on the lattice spacing $a$ of the relevant (dimensional) observables.
Let us study first the symmetric case
$\beta_s=\beta_t\equiv\beta\equiv\frac{4}{g^2}$.
 Then the dependence of the lattice spacing on $\beta$  is known in the
continuum limit in form of the renormalization group equation:
\eq
a\Lambda_L=\left({b_0 g^2}\right)^{-\frac{b_1}{2b_0^2}}
\exp\left(-\frac{1}{2b_0g^2}\right)~~,
\label{lambda}
\en
where $\Lambda_L$ is the lattice scale parameter (in units of
which we must measure any dimensional quantity on the lattice) and $b_0, b_1$
are the first two coefficients of the Callan-Symanzik equation:
\eq
b_0=\frac{11 N}{48\pi^2},\hskip 1cm
b_1=\frac{34}{3}\left(\frac{ N}{16\pi^2}\right)^2~~~,
\en
with $N=2$ in our case.

Here and in the following we have fixed the spacetime dimensions to be
(3+1). This is a
particularly important remark since it is only in (3+1) dimensions that the
coupling constant $\beta$ is adimensional and the renormalization group 
equations
have this peculiar exponential behaviour.

Plugging eq.(\ref{lambda}) into the definition of critical temperature:
\eq
T_c=\frac{1}{aN_t}
\label{tcsym}
\en
we find:
\eq
\frac{T_c}{\Lambda_L}=\frac{1}{N_t}
\left(\frac{6\pi^2\beta}{11}\right)^{-\frac{51}{121}}
\exp\left(\frac{3\pi^2\beta}{11}\right) .
\label{tccont}
\en

If the continuum limit is correctly reached then the ratio $T_c/\Lambda_L$
should approach for large enough values of $\beta$ (hence, in our case, also for
large values of $N_t$) a constant value. Plugging this constant into eq.
(\ref{tccont}) we immediately recover the well known (approximate) logarithmic
growth of  $\beta_c$ as a function of $N_t$, which is the typical signature
of the correct continuum limit behaviour of the deconfinement
temperature in a (3+1) dimensional LGT.
It is important at this point to stress that this logarithmic behaviour 
 is a  non trivial requirement for any effective theory
approach to the deconfinement transition. For instance,
in the zeroth order approximation of the effective action, the effective
coupling constant is
written as a combination of modified Bessel functions raised to the $N_t$
power (see eq.(\ref{a2})). Since the  large $\beta$
asymptotic behaviour of the Bessel function $I_n(\beta)$ is
\eq
I_n(\beta)
\sim\frac{e^\beta}{\sqrt{2\pi\beta}}\left[1-\frac{4n^2-1}{8\beta}
+\cdots\right]
\label{asybes}
\en
then the effective couplings scales as:
\eq
\left[\frac{I_n(\beta)}{I_1(\beta)}\right]^{N_t}
\sim 1-\frac{(n^2-1)N_t}{2\beta}
+\cdots ,
\label{asybes2}
\en
which implies a {\sl linear} scaling of $\beta_c$ as a function of $N_t$.

The lack of logarithmic scaling in the zeroth order approximation is
one of the main reasons which motivated 
us to look at higher order corrections in the effective action.

\subsection{Asymmetric Lattices}

The fact that we have in general asymmetric couplings $\beta_s\neq\beta_t$
adds some further complication to the previous discussion, but has some very
important consequences. 
For each value of $\rho$ we have a new independent
regularization scheme, with an independent renormalization group equation. 
This means that, if we define the coupling $g$ according to eq.(\ref{couplings})
as 
$\frac{4}{g^2}=\sqrt{\beta_s\beta_t}$, we must substitute eq.(\ref{lambda})
with:
\eq
a\Lambda(\rho)=\left({b_0 g^2}\right)^{-\frac{b_1}{2b_0^2}}
\exp\left(-\frac{1}{2b_0g^2}\right)~~.
\label{lambdarho}
\en
In other words we have that now the $\Lambda$ parameter is also a 
function of $\rho$. The last step, in order to obtain meaningful results in the
continuum limit is then to relate the  scale $\Lambda(\rho)$ to the $\Lambda$
parameter of the symmetric lattice regularization. This problem was studied
in~\cite{k81}
where the ratio
$\Lambda(\rho)/\Lambda_L$ (with $\Lambda_L\equiv\Lambda(1)$ denoting the
scale of the symmetric case
$\beta_s=\beta_t$ discussed above) was evaluated explicitly at one loop.
It was found to be a universal
function of $\rho$:
\eq
\frac{\Lambda(\rho)}{\Lambda_L}=
\exp\left(-\frac{c_\sigma(\rho)+c_\tau(\rho)}{4b_0}\right) ,
\label{lrl}
\en
where $c_\sigma$ and $c_\tau$ are the functions already introduced in 
eq.s (\ref{rel3}) and (\ref{rel4}).
It is easy to see that this result is completely equivalent to the 
shift in $\beta$ of eq.s
(\ref{rel3}) and (\ref{rel4}). 
It gives us a tool to better understand eq.(\ref{rel3}), which
simply encodes the effect of the quantum fluctuations at one loop.

This result is particularly interesting for our purposes, since it allows to
extend our analysis also to asymmetric lattices, taking into account also the
quantum effects. 
Let us discuss in detail this point.

Consider an asymmetric regularization, with
$\beta_t=\rho^2\beta_s$, $\rho>1$, defined on a lattice with
temporal extension $\widetilde N_t$.
By using eq.(\ref{rel1}) and (\ref{rel2})
 we see that this regularization  is (classically) equivalent
to that on a symmetric lattice with $\beta\equiv \frac{\beta_t}{\rho}
=\beta_s\rho$ and with temporal extension $N_t = {\widetilde N_t \over \rho}$.
 Hence $\rho$  must be a rational
number; we shall choose in general $\rho$ to be an integer number, so that
$\widetilde N_t$ will be a multiple of $N_t$. 
At the quantum level we can still show the
equivalence of the symmetric and asymmetric lattice regularizations provided we modify
the previous relations according to eq.(\ref{rel3})
and (\ref{rel4}).
By using the explicit knowledge of the $\rho\to\infty$ expansion of 
$c_{\sigma}$ and $c_{\tau}$, we obtain, 
for large enough values of $\rho$, the following
scaling behaviour:
\eq
\beta_{t,c}(\rho)=(\beta_c+\alpha^0_t)\rho+ \alpha^1_t
\label{rho3}
\en
where $\beta_{t,c}(\rho)$ is the critical coupling on the asymmetric lattice
and $\beta_c$ the critical coupling on the equivalent symmetric lattice.
The numerical values of $\alpha^0_t$ and $\alpha^1_t$ are reported in sect. 2,
following eq.(\ref{rhoex}).  
Higher order corrections to eq.(\ref{rho3}) vanish as $\rho\to \infty$. 

The limit $\rho\to\infty$ is particularly interesting because some
relevant simplifications occur in that limit in the effective action.
Let us consider first the contributions of the timelike plaquettes in 
the zeroth order approximation given in eq.(\ref{a2}).  
We have now to perform in (\ref{a2}) the following substitution:
\eq
 N_t  \to  {\widetilde N_t} = \rho N_t ~~~~,~~~~~\beta_t  \to \beta_t(\rho)
\label{subs}
\en
with $\beta_t(\rho)$  given in eq.(\ref{rho3}), and finally take 
the limit $\rho\to\infty$ . 
By using the asymptotic expansion of the Bessel functions (\ref{asybes}) one 
obtains in that limit that the effective couplings at the r.h.s. of eq.(\ref{a2})
become
\eq
\left[\frac{I_{2 j+1}(\beta(\rho))}{I_1(\beta(\rho))}\right]^{\widetilde N_t}
\hskip 10pt\stackrel{\rho\to\infty}{\sim}\hskip 10pt
\exp (-{2 j (j+1) N_t \over \beta + \alpha^0_t}) ,
\label{htker}
\en
where $\beta $ is the coupling of the corresponding symmetric lattice. 
One can easily recognize  the quadratic Casimir of the representation  $j$
at the exponent, and one immediately 
realizes that the action for the time-like plaquettes becomes the Heat-Kernel 
action in the $\rho\to\infty$ limit, namely in the hamiltonian limit. 

The couplings of the spacelike plaquette in the adjoint representation 
and of the pair of plaquettes in the fundamental representation are 
respectively proportional to 
$3 {\widetilde N_t} {I_3(\beta_s(\rho)) \over I_1(\beta_s(\rho))}$ and $ 4
{\widetilde N_t}^2 ({I_2(\beta_s(\rho)) \over I_1(\beta_s(\rho))})^2$ 
(see eq.(\ref{act3})).
We replace ${\widetilde N_t}$ as in (\ref{subs}) and $\beta_s(\rho)$ by
${\beta + \alpha^0_s \over \rho}$ and take the limit $\rho\to\infty$.
The contribution of the adjoint 
representation is of order $1/\rho$ and hence it vanishes in this limit, due 
essentially to its zero measure in an infinite lattice. On the contrary the
coupling of the pair of plaquettes in the fundamental representation is finite
in the limit $\rho\to\infty$ and its limiting value is
$N_t^2 (\beta +  \alpha^0_s)^2 /4$. Notice that all  powers of 
$\beta_s(\rho)$ higher than $\beta_s(\rho)^2$ in the power expansion of the 
Bessel functions give a vanishing contribution in 
the limit $\rho\to\infty$. 
This means that, as already remarked at the beginning 
of sect. 3, the character expansion and the power 
expansion coincide in this limit, thus removing any possible ambiguity.
Notice also that in spite of the vanishing of $\beta_s(\rho)$ in the 
asymmetric limit the effective expansion parameter 
$N_t (\beta +  \alpha^0_s)$ is never
very small and it does indeed increase as we approach the continuum limit.

\subsection{Results.}
In order to extract reliable estimates for the critical
couplings, we must first discuss the behaviour of our results as functions of
the asymmetry parameter $\rho$.
\subsubsection{$\rho$ dependence of the results}

It is very interesting to study the $\rho$ dependence of our results, to see if
eq.(\ref{rho3}) is fulfilled.
\begin{table}
\label{tab1}
\begin{center}
\begin{tabular}{c c c c c c}
\hline\hline
$\rho$ & $\widetilde N_t$ & $\beta_{t,c}$ & $\beta_{s,c}$
& $\delta$ & $\gamma$ \\
\hline
$1$ & $6$ & $3.216$ & $3.216$ & $$ & $$\\
$2$ & $12$ & $5.882$ & $1.471$ & $2.666$ & $0.55$\\
$3$ & $18$ & $8.576$ & $0.953$ & $2.694$ & $0.49$\\
$4$ & $24$ & $11.282$ & $0.705$ & $2.706$ & $0.46$\\
$5$ & $30$ & $13.993$ & $0.560$ & $2.711$ & $0.45$\\
$6$ & $36$ & $16.707$ & $0.464$ & $2.714$ & $0.42$\\
$7$ & $42$ & $19.422$ & $0.396$ & $2.715$ & $0.42$\\
$8$ & $48$ & $22.138$ & $0.346$ & $2.716$ & $0.41$\\
$10$ & $60$ & $27.571$ & $0.276$ & $2.717$ & $0.39$\\
$20$ & $120$ & $54.753$ & $0.137$ & $2.718$ & $0.39$\\
\hline\hline
\end{tabular}
\end{center}
\vskip 0.3cm
{\bf Tab. I}{\it~~ The critical coupling $\beta_{t,c}$ as a function of
$\rho$. In the second column we have reported the temporal extension
$\widetilde N_t =\rho N_t$ of the asymmetric lattices that we used. In the fourth
column we have reported the values of $\beta_{s,c}\equiv\beta_{t,c}/\rho^2$.
In the last two
columns we have reported the values of $\gamma$ and $\delta$ (see text for
explanation). The values of $\delta$ reported in Tab. I are obtained by 
applying the definition (\ref{delta}) with $\rho_2=\rho$ and $\rho_1$ equal
to the value of $\rho$ in the previous row.}
\end{table}
In Tab. I we have reported, as an example, the $\rho$ behaviour of a set of
asymmetric regularizations which are all equivalent to the symmetric $N_t=6$
case. For each pair of (subsequent) values of $\beta_{t,c}(\rho)$ we have
constructed the two quantities
\eq
\delta=\frac{\beta_{t,c}(\rho_2)-\beta_{t,c}(\rho_1)}{\rho_2-\rho_1}
\hskip 2cm (\rho_2>\rho_1)
\label{delta}
\en
\eq
\gamma=\beta_{t,c}(\rho_2)-\delta\rho_2
\en
which, if eq.(\ref{rho3}) is fulfilled, 
 should provide a good estimate of  $\beta_c+\alpha^0_t$ and $\alpha^1_t$
respectively. Their values together with those of
$\beta_{t,c}$ are reported in Tab. I.

It can be seen that the data follow very well the expected law. In particular
it is clearly visible the $1/\rho$ quantum correction, which is definitely
different from zero and smoothly approaches for large values of 
$\rho$ the value $\gamma\sim 0.39$, which is not too far from the 
expected value  $\alpha_t^1=0.5$.

The values of $\beta_{s,c}\equiv\beta_{t,c}/\rho^2$, that we have reported in
the fourth column of Tab. I, give an idea of the reliability of our
 strong coupling expansion in $\beta_s$.

We repeated the same analysis for all the values of $N_t$ for which Montecarlo
data are known (the MC data are reported in Tab. IV). Our
results are collected in Tab. II where we have reported the asymptotic (large
$\rho$) values of $\gamma$ and $\delta$ as well as of another 
quantity $\epsilon$ that should provide us with an estimate of
$\alpha^0_t$. This is defined as $\epsilon = \delta - \beta_c$,
where $\beta_c$ is the critical coupling obtained with a symmetric lattice of 
size $N_t$ by keeping strictly only the terms of order $\beta_s^2$ in the
character expansion. In other words, in computing $\beta_s$  we neglect
all contributions that vanish in the limit $\rho \to \infty $, thus keeping 
the same contributions in the symmetric and asymmetric lattice. 
Tab.II shows that, in the range $N_t=2-16$,
the agreement with the theoretically expected values of $\alpha_t^0$ and
$\alpha_t^1$ is really remarkable. Let us stress again that this agreement is
highly non trivial since $\alpha_t^0$ and $\alpha_t^1$ were obtained with a
{\sl weak coupling} calculation, while our effective action is the result of a
{\sl strong coupling} expansion. The reason of this success is very likely related
to the fact that we have been able to sum to all orders in $\beta_t$
 the timelike contribution of the effective action.

\begin{table}
\label{tab2}
\begin{center}
\begin{tabular}{c c c c }
\hline\hline
$ N_t$ & $\delta$& $\epsilon$ & $\gamma$ \\
\hline
$2$ &  $1.554$ & $-0.184$  & $0.414$\\
$3$ &  $1.971$ & $-0.210$  & $0.375$\\
$4$ &  $2.259$ & $-0.221$  & $0.373$\\
$5$ &  $2.500$ & $-0.235$  & $0.372$\\
$6$ &  $2.718$ & $-0.249$  & $0.389$\\
$8$ &  $3.114$ & $-0.271$  & $0.413$\\
$16$ & $4.443$ & $-0.327$  & $0.508$\\
\hline
     &         & $-0.27192$ & $0.50$\\
\hline\hline
\end{tabular}
\end{center}
\vskip 0.3cm
{\bf Tab. II}{\it~~ Values of $\delta$, $\epsilon$ and $\gamma$ as functions
of $N_t$ (see the text for the definitions of these three quantities).
$\epsilon$ and $\gamma$ are estimators of $\alpha_t^0$ and $\alpha_t^1$ whose
theoretical values are reported, for comparison, in the last row of the 
table.}
\end{table}

\subsubsection{Scaling behaviour.}

The agreement between the $\rho$ dependence of our mean field results and the
theoretical expectations allows us to be  confident on their consistency, 
also at the quantum level. So in order to extract our best
estimate for the critical coupling we can consistently  
{\sl assume} eq.s (\ref{rel3}) and
(\ref{rel4}), with the correct theoretical $\alpha$ coefficients, work in
the large $\rho$ limit (so as to avoid ambiguities in the $\beta_s$ expansion)
and then rescale the resulting critical values to the limit of symmetric
lattice\footnote{Let us notice, as a side remark, that 
the difference between the values of $\beta_c$ obtained in this way and those
which one would obtain with the naive procedure of choosing right from the
beginning a symmetric lattice is not very large, but nevertheless it is not
negligible and could become very important if further orders in the $\beta_s$
expansion were to be added to the effective action.}, so as to allow a 
comparison with the Montecarlo results.

The values obtained in this way for the critical couplings
 are reported in  Tab. III, and plotted in
Fig. 2, where they are also compared with the Montecarlo results
(extracted from~\cite{fhk}), which are reported in Tab. IV .
\iffigs
\begin{figure}
\null\vskip -1cm\hskip 2cm
\epsfxsize = 12 truecm
\epsffile{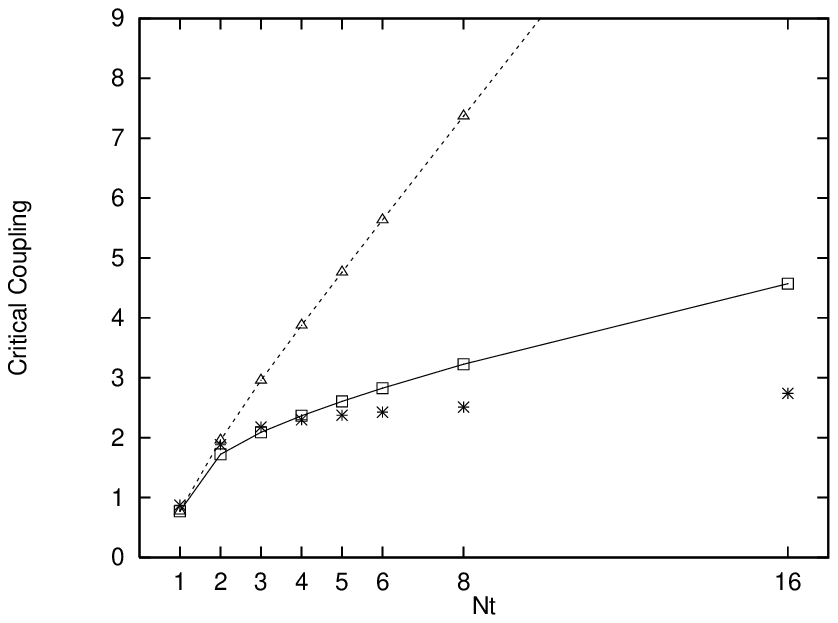}
\vskip 0.3cm
{\bf Fig. 2} {\it Values of the critical coupling $\beta_c$ are plotted for
different values of the number of time-like links $N_t$. Results obtained
with Montecarlo simulations, which are denoted by *,
 are compared with those obtained with our
mean field analysis: $\triangle$ represents the data for $\beta_{c}|_{0}$,
the critical coupling in the zeroth order approximation and $\Box$  
the data for $\beta_c|_{1,\rho\to\infty}$, 
the critical coupling including the effect of
the space-like plaquettes al the lowest non trivial order, calculated in the
limit $\rho \to \infty$ and reduced to the value $\rho=1$ as described in the
text.}
\label{data}
\end{figure}
\fi
It is impressive to see how the contribution of the space-like plaquettes,
although taken into account at the order  $\beta_s^2$ only, 
improves the agreement of the results
of the mean field method with the ones of the Montecarlo simulations.
To the order $\beta_s^2$ the mean field method gives results which are 
in reasonably good agreement with the Montecarlo for $N_t$'s as high
as 5, displaying in the range $ 1 \leq N_t \leq 5 $ a behaviour which is
compatible with the logarithmic scaling predicted by the renormalization 
group. In contrast the model with only time-like plaquettes shows from
$N_t=1$ a linear scaling behaviour in contradiction with both renormalization
group predictions and Montecarlo simulations.
It must be  stressed  however that the almost perfect agreement of our results
with the Montecarlo simulations for small $N_t$'s (especially for $N_t $ in the
range $(3-4)$ ) is most probably only apparent, resulting from a compensation
between an expected 10-15\% systematic error due
to the mean field approximation and the effects of higher orders contributions. 
\begin{table}
\label{tab3}
\begin{center}
\begin{tabular}{c c c}\hline\hline
$N_t$ & $\beta_c\vert_0$ & 
$\beta_c\vert_{1,\rho\to\infty}$ \\
\hline
$2$ & $1.957$ & $1.723$  \\
$3$ & $2.957$ & $2.089$  \\
$4$ & $3.876$ & $2.366$  \\
$5$ & $4.763$ & $2.606$  \\
$6$ & $5.636$ & $2.826$  \\
$8$ & $7.369$ & $3.227$  \\
$16$ & $14.262$ & $4.572$ \\
\hline\hline
\end{tabular}
\end{center}
\vskip 0.3cm
{\bf Tab. III}{\it~~ The critical coupling $\beta_c$
as a function of the
lattice size $N_t$ in the t direction. In the second column the values
obtained with the zeroth order approximation and in the third column those
obtained at the order $\beta_s^2$ by rescaling 
the values obtained in highly asymmetric lattices $\rho \to\infty$ as
explained in the text. }
\end{table}
\begin{table}
\label{tab4}
\begin{center}
\begin{tabular}{c c c}
\hline\hline
$N_t$ & $\beta_c$ & $T_c/\Lambda$ \\
\hline
$2$ & $1.8800(30)$ & $29.7(2)$ \\
$3$ & $2.1768(30)$ & $41.4(3)$ \\
$4$ & $2.2986(6)$ & $42.1(1)$ \\
$5$ & $2.3726(45)$ & $40.6(5)$ \\
$6$ & $2.4265(30)$ & $38.7(3)$ \\
$8$ & $2.5115(40)$ & $36.0(4)$ \\
$16$ & $2.7395(100)$ & $32.0(8)$ \\
\hline\hline
\end{tabular}
\end{center}
\vskip 0.3cm
{\bf Tab. IV}{\it~~ The critical coupling $\beta_c$ and the
corresponding deconfinement temperature $T_c/\Lambda$
as a function of the
lattice size in the t direction, $N_t$, in the (3+1) dimensional SU(2)
LGT. The data are taken from~\cite{fhk}.}
\end{table}

\subsection{$N_t=1$ again: improving the mean field approximation.}
As we remarked in sect. 3.5, the most interesting feature of 
the $N_t=1$ case is that, due to the
remarkable simplifications  which occur in this case, the character 
expansion
of the  effective action can
be resummed exactly. This allows a much simpler implementation of improved
versions of the mean field approximation, and we decided to use this case as
a laboratory to test these improved estimators and above all to have a hint of
the magnitude of the systematic errors involved in the plain mean field
approximation that we used in the previous section.
Notice also that for this particular $N_t=1$ case
it does exist a very precise Montecarlo estimate of the critical coupling,
namely $\beta_c=0.8730(2)$~\cite{bems}, 
and we shall use this value as a reference
point to compare our predictions.

Let us notice, as a preliminary remark, that in the $N_t=1$ case the
$\beta^2_s$ contribution is given by the adjoint plaquette term only, since
there is no room to locate a pair of  plaquettes in the fundamental
representation. This fact has two consequences: first, the magnitude of the
correction to the critical temperature due to the spacelike contribution is
much smaller than in the $N_t>1$ case; second, it has the opposite sign, namely
$\beta_{t,c}$ {\sl increases} as a consequence of the spacelike term.

The simplest way to improve the mean field approximation is to consider
larger and larger clusters of spins (see Fig. 3). It can be shown that this
modification indeed allows a more and more precise determination of the critical
coupling. The first improvement (step 2 in the notation of Fig. 3) is also
known as Bethe approximation \cite{book}. 
In Tab. V we report the results of our analysis.
It is easy to see that both with and without the spacelike plaquettes
the plain mean field approximation is affected by an error which ranges from
10\% to 15\% (depending on the role which is played by the spacelike
contribution in the Montecarlo estimate), and that almost half of this gap is
filled by the Bethe approximation.

We will show in a forthcoming publication \cite{noi2}, that  
the results of next order approximation (step 3 in
the notation of Fig. 3) 
differ from the exact result by only a few per cent.
Notice however that the step 3 approximation is very hard to handle and
requires several technical manipulations. 
In this forthcoming paper, we will also show that
a result which is very close to the Montecarlo one given in last 
column of Tab. V
can be obtained by comparing the different steps of Bethe approximations
in the present model with the corresponding ones in the Ising model, whose
critical coupling is known with high precision.  

\iffigs
\begin{figure}
\null\vskip -1cm\hskip 2cm
\epsfxsize = 12 truecm
\epsffile{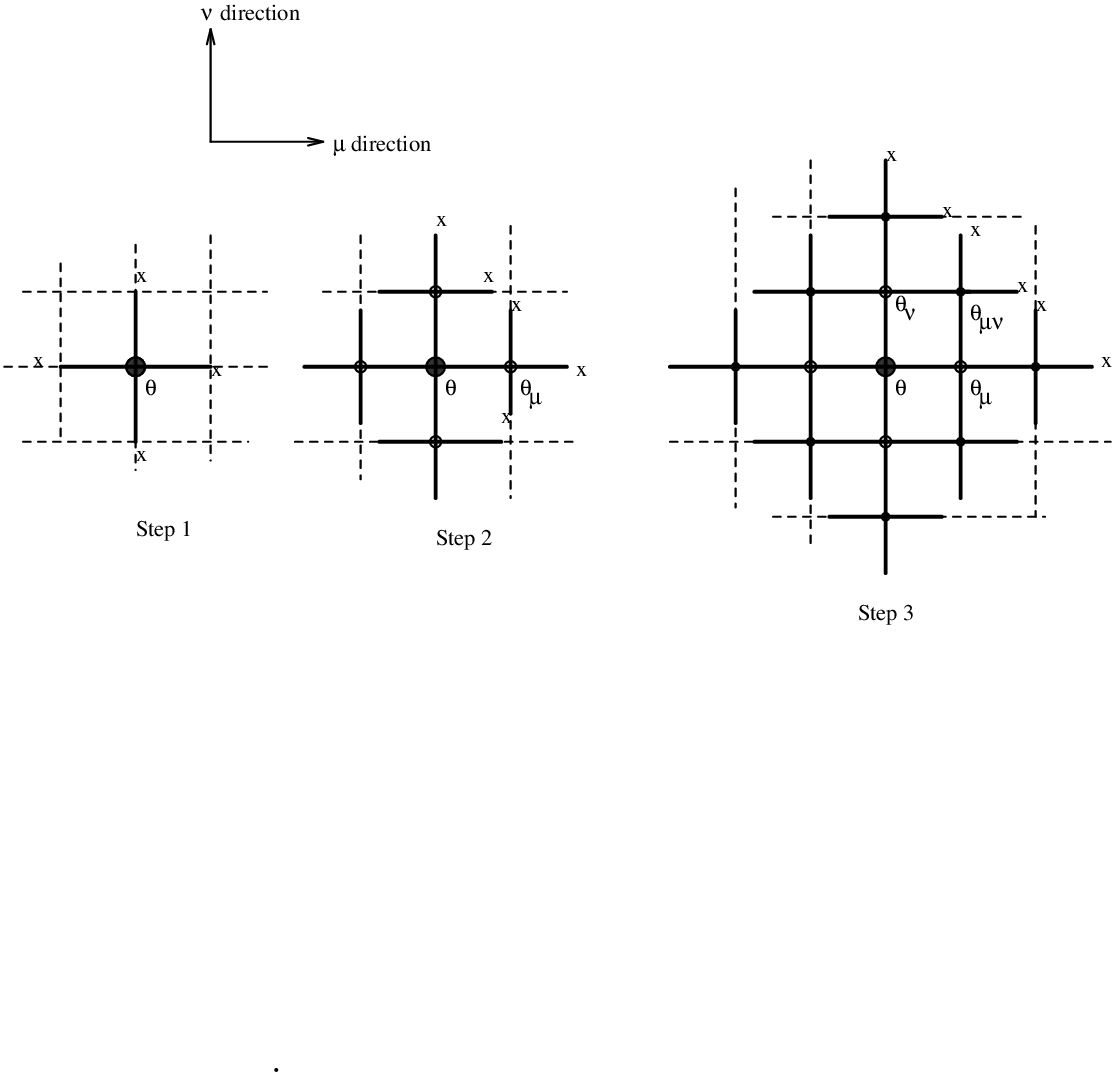}
\vskip -4cm
{\bf Fig. 3} {\it Progressively more precise Bethe-like
approximations}
\label{contourplots}
\end{figure}
\fi
\begin{table}[htb]
\label{tab5}
\begin{center}
\begin{tabular}{c c c}
\hline\hline
step 1 & step 2 & MonteCarlo \\
\hline
$0.7702$ & $0.8139$  & $ - $\\
$0.7705$ & $0.8145$  & $0.8730(2)$\\
\hline\hline
\end{tabular}
\end{center}
\vskip 0.3cm
{\bf Tab. V}{\it~~ The critical coupling $\beta_c$ for the action
$S_0$ (first row) and $S_0+S_1$ (second row)
in the mean field approximation (step 1) and the Bethe approximation
(step 2) . In the last column the
Montecarlo result of ref.~\cite{bems} (including full contribution of
space-like plaquettes).}
\end {table}
An alternative approach one can follow is to identify explicitly
the Ising model which is hidden in the
effective action (following  the Svetitsky--Yaffe analysis) and then use again 
our knowledge of the critical coupling of the three dimensional Ising model. 
This approach has been developed for the $N_t = 1$ case in~\cite{var} and 
it will
also be fully exploited in~\cite{noi2}.

\section{Conclusions}
The main result of our paper is the construction of the effective
action for the Polyakov loops at the first non trivial order in the spacelike
coupling, which is exact to all orders in the timelike coupling.

Extracting the critical temperature from this effective action, and comparing
our results with those of the Montecarlo simulation, we have seen that the
effective action describes rather well
(even within the mean field approximation)  the full theory up to a lattice size in
the compactified time direction of $N_t\sim 5$. This is an impressive
improvement with respect to the previous studies, in which
the spacelike contributions were neglected and which were constrained to
$N_t=1$. We see in principle no obstruction to extend our analysis to higher
orders and to reach a better and better agreement with the renormalization group
expectations. Such a result would obviously be very interesting, since it would
be the first time that one can reach the scaling regime of a dimensional
quantity (as the critical temperature) by using only analytical tools.
{}From this point of view the critical temperature seems indeed in a better
position with respect for instance to the string tension, for  which, due to the
roughening transition, a strong coupling approach has very few chances to reach
the scaling region.
It is also remarkable the agreement between our results, based essentially on
a strong coupling expansion, and the weak coupling calculations of Karsch, 
regarding the consistency of the theory with respect to lattice deformations 
described by the asymmetry parameter $\rho$. 

Moreover our approach can be straightforwardly extended to the SU(3) gauge 
model,
which is obviously more interesting from a phenomenological point of view.

Besides these obvious remarks there are two more reasons of interest of our
result, which we think should deserve some attention. First, it was noticed
in~\cite{fhk} that the ratio $T_c/\sqrt{\sigma}$ shows a very precocious
scaling, and is essentially stable already for lattices as small as $N_t=4$.
This value lies inside the region where we have seen we can
trust  our expansion. Therefore,
if one were able to extract the string tension in our framework, it would be
possible to reach the continuum limit value of $T_c/\sqrt{\sigma}$ already 
within our (relatively simple) effective action.

A second possible interesting application is in the study of LGT with suitable
generalizations of the Wilson action.  
In particular,  much interest has been recently
devoted to  the so called fundamental-adjoint SU(2) LGT (see~\cite{mg}
and references therein) in which a term proportional to the plaquette in the
adjoint representation is added to the ordinary Wilson action. This introduces
a new coupling $\beta_A$. In~\cite{mg} the
behaviour of the deconfinement transition in the extended
coupling plane ($\beta,\beta_A$) was studied with a strong coupling effective
action truncated to the first order and for $N_t=2$ and
$N_t=4$, hence in a region in which our approach seems to have a good
behaviour. Since our action is written in terms of a character
expansion, it should be rather straightforward to generalize it to include
a coupling  $\beta_A$ different from zero.

\vskip 1cm
\centerline{{\bf  Acknowledgements}}
\vskip 0.5cm
We thank P. De Los Rios, F. Franjic, G. Gionti for many helpful discussions.

\end{document}